\documentclass[12pt,preprint]{aastex}
\usepackage{graphicx}
\newcommand{\kms}{km~s$^{-1}$}
\newcommand{\smpy}{$M_{\sun}\ yr^{-1}$}
\newcommand{\etal}{et al.}
\begin{document}

\title{Fast Winds and Mass Loss  from Metal-Poor Field 
Giants\footnote{Data presented herein were 
 obtained at the W. M. Keck 
 Observatory, which is operated as a scientific partnership among the 
California Institute of Technology, the University of California, and the 
 National Aeronautics and Space Administration. The Observatory was made
 possible by the generous financial support of the W. M. Keck Foundation}}

\author{A. K. Dupree}  
\affil{Harvard-Smithsonian Center for Astrophysics, 60 Garden Street,
  Cambridge, MA 02138}
 \email{dupree@cfa.harvard.edu}

\author{Graeme H. Smith}
\affil{University of California Observatories/Lick Observatory, 
 University of California, Santa Cruz, CA 95064}
\email{graeme@ucolick.org}

\author{Jay Strader\footnote{Hubble Fellow}}
\affil{Harvard-Smithsonian Center for Astrophysics, 60 Garden Street,
  Cambridge, MA 02138}
\email{jstrader@cfa.harvard.edu}

\begin{abstract}

Echelle  spectra  of the infrared \ion{He}{1} $\lambda$10830 line were obtained 
with NIRSPEC on the Keck 2 telescope for 
41 metal-deficient  field giant stars including those on the red giant
branch (RGB), asymptotic giant branch (AGB), and red  horizontal
branch (RHB).  The presence of this \ion{He}{1} line is ubiquitous 
in  stars with $T_{eff} \gtrsim$ 4500K and $M_V$ fainter than $-$1.5, 
and reveals the dynamics of the atmosphere. The line
strength  increases with effective 
temperature for $T_{eff} \gtrsim$ 5300K in RHB stars.
In AGB and RGB stars, the line strength
increases with luminosity.   Fast outflows ($\ga$ 60 km s$^{-1}$) are detected 
from the majority of the 
stars and about 40 percent of the outflows have sufficient 
speed as to allow escape of material from 
the star as well as
from a globular cluster. Outflow speeds and line strengths do not 
depend on metallicity for our sample ([Fe/H]= $-$0.7 to $-$3.0) 
suggesting the driving mechanism for these
winds derives from  magnetic and/or hydrodynamic processes. 
Gas outflows are present in every luminous giant, but are not detected
in all stars of lower luminosity indicating possible variability.
Mass loss rates ranging 
from $\sim 3 \times 10^{-10}$ to $\sim 6 \times 10^{-8}$ \smpy\  estimated
from the Sobolev approximation for line formation 
represent values with  evolutionary significance for red giants and red
horizontal branch stars. We estimate that 0.2 M$_\sun$ will be lost on
the red giant branch, and the torque of this wind can account for 
observations of slowly rotating RHB stars in the field.  About 0.1--0.2 M$_\sun$ 
will be lost on the red horizontal branch itself.
This first  empirical determination of mass loss on the RHB may
contribute to the appearance of extended horizontal branches in
globular clusters.  The spectra appear to resolve the problem of
missing intracluster material in globular clusters.  Opportunities 
exist for 'wind smothering' of
dwarf stars  by winds from the evolved population, possibly leading to
surface pollution in regions of high stellar density. 

\end{abstract}

\keywords{stars: chromospheres --- stars: Population II --- stars: winds, 
outflows}

\section{Introduction}
The assumption of mass loss from stars evolving on the red giant
branch of globular clusters has yet to be tested  through direct
detection of winds.  This assumption remains one
of the major concerns in the evolution of low mass stars and may
be related to  the 
second-parameter problem (Sandage \& Wildey 1967): differing 
horizontal branch morphology between globular clusters of the 
same metallicity and age. Various explanations have been offered for 
differences in the horizontal branch morphology including intrinsic 
dispersions in the amount of stellar mass loss, rotation, or 
deep mixing, environmental 
effects possibly correlated with cluster mass or central density, 
heterogeneities in He abundance possibly as a result of cluster pollution by
intermediate-mass asymptotic giant branch (AGB) stars, or of the infall of
planets onto cluster stars (Buonanno et al. 1993; Buonanno et al. 1998; 
Catelan et al. 2001; Sills \& Pinsonneault 2000; Recio-Blanco et al. 2006;
Sandquist \& Martel 2007; Soker et al. 2001b; Sneden et al. 2004; 
Sweigart 1997; Peterson et al. 1995;  Ventura \& D'Antona 2005).   
Most recently 
with the advent of infrared photometry discussed below,
and the identification of multiple populations
on the main sequence of  globular clusters with the {\it Hubble Space
  Telescope} (Anderson 2002; Bedin et al. 2004; Piotto et al. 2007),
renewed attention is focussing on the mass loss process.

A related issue is the puzzling absence of the material
lost from the red giants in a globular cluster (Tayler \& Wood 1975).
This gas is expected to accumulate in clusters between sweeps
through the galactic plane, but detection has proved elusive.  Unconfirmed
measures of 21-cm H-line emission exist for NGC 2808 (Faulkner et
al. 1991). The radio dispersion of milli-second pulsars
located in 47 Tuc and M15 hinted at an enhanced electron
density in the intracluster medium (Freire et al. 2001). Deep
searches for H I in 5 clusters gave one firm detection of \ion{H}{1}
emission, and possibly 2 others, but the amount of mass inferred from 
this is a few orders of magnitude less than expected (van Loon et
al. 2006).  A search for intracluster dust in 12 globular clusters
with the Far-Infrared Surveyor on {\it AKARI} failed to detect
emission except possibly in one cluster (Matsunaga et al. 2008). 
There is a notable lack of detections of intracluster
material with the {\it Spitzer Space Telescope}. Only 2 clusters of the many
observed with {\it Spitzer} have 'possible' detections of
intracluster material (Boyer \etal\ 2006; Barmby \etal\ 2009).
One suggestion for removal of intracluster material
is ram-pressure stripping by the galactic halo 
(Frank \& Gisler 1976, Okada \etal\ 2007),
but {\it Spitzer} observations can not confirm the expected relation
between cluster kinematics and the presence (or upper limit) of dust (Barmby
\etal\ 2009).

Results from mid-infrared observations of metal-deficient field giants
and globular clusters with  {\it IRAS}, the {\it ISO} satellite,  
the {\it Spitzer Space Telescope}, and {\it AKARI} 
(Smith 1998; Origlia et al. 2002, 2007; Evans et al. 2003; 
Boyer et al. 2006; Ita et al. 2007; Boyer et al. 2008)  suggest 
that some giants  have produced circumstellar
dust that could result from   stellar winds.  Since not all
giants display excess infrared emission, mass loss associated with
dust appears to be 
episodic (Origlia et al. 2007;  M\'esz\'aros et al. 2008).
Caloi and D'Antona (2008) further suggest that the mass loss rate might
be `sharply' peaked at one  value along the red giant branch.  
In the metal-rich open cluster NGC 6791, 
the presence of low mass white dwarfs  led Kalirai 
et al. (2007) to conclude that mass loss is enhanced in high
metallicity environments, although that suggestion appears
to be controversial (van Loon et al. 2008; Bedin et al. 2008). 
Concurrently with the above results, the possibility of multiple 
episodes of star-formation, and
self-pollution in globular clusters is receiving increased attention 
as a means to explain
chemical variations and multiple branches in the color-magnitude 
diagram of clusters
(Lee \etal\ 1999; Pancino \etal\ 2000; Anderson 2002; D'Antona \etal\ 2002; 
Piotto \etal\ 2007; Kayser \etal\ 2008). This suggestion frequently 
resorts to AGB stars with mass $>~$3M$_\sun$ `polluting' either the surface of 
the cluster stars or the environment  in which a second
generation of stars subsequently forms.  The wind velocities 
from the AGB stars must
be slow so that material will not  escape the cluster.  
A long-standing suggestion envisions that pulsation near the
top of the AGB may degenerate into relaxation oscillations, during
which mass loss or envelope ejection occurs in rapid fashion, dubbed
a `superwind' (Renzini 1981; Bowen and Willson 1991).  Another
conjecture (Soker et al. 2001a) calls for a superwind during the immediate post-RGB
phase to explain gaps on the horizontal branch.  Clearly,
both observations and theory currently allow a great variety  in both 
the presence and character of mass loss from cluster stars.  The
spectra
of \ion{He}{1} $\lambda$10830 presented here can address these questions.

\section{Spectroscopic Diagnostics of Winds}

Spectra of stars contain many features that can be used to detect
winds directly, but their diagnostic properties are related to the specific conditions in  
a stellar atmosphere. In a cool giant star, atoms and ions  form
at different levels in the chromosphere, loosely tied to the local electron temperature
which  is increasing  with height.  Additionally, for strong 
lines such as H$\alpha$, \ion{Ca}{2} H and K, and \ion{Mg}{2}, 
the profile  is formed over an extended range of atmospheric layers which
results from differing opacities across the line itself. Calculations for metal-poor 
giants  demonstrate (Dupree et al. 1992a; Mauas et al. 2006) the 
locations in the atmosphere where commonly observed lines are 
formed.  Mass flows can produce asymmetric profiles of H$\alpha$ and Na
D, as well as emission asymmetries and velocity shifts in the reversed
absorption core of \ion{Ca}{2} and \ion{Mg}{2}. These features    
have been measured in red giants in many globular clusters 
(Cacciari et al. 2004; McDonald and van Loon 2007; 
M\'esz\'aros et al. 2008, 2009b) and in metal-poor field stars (Smith \&
Dupree 1988; Dupree \& Smith 1995; Dupree et al. 2007). However the 
velocities inferred from the optical profiles are generally less than the 
escape velocity from the star, and  can not truly be identified as
stellar winds.

We note the difference between the use  of 'outflow' and 'wind' 
in this context. By 'outflow', we mean the presence of a velocity
field within the region of a chromosphere sampled by our spectral 
diagnostic in which a flow of material occurs that is  moving 
away from the stellar photosphere.  The term 'wind' is reserved for
the particular case of an outflow in which the outflow velocity
exceeds the escape velocity within the region of the chromosphere
sampled by our spectral diagnostic.

A good diagnostic of winds is the near-infrared He I 10830\AA\ line 
($1s2s\ ^3S\ -\ 1s2p\ ^3P$) which models
show (Dupree et al. 1992a) is formed higher in the metal-poor atmosphere than 
H$\alpha$ and \ion{Ca}{2} K.  Thus it might be expected to trace out
higher velocities, where the outflow becomes a wind, than the  optical diagnostics.
Additionally, the lower level of this transition is metastable, and 
is not closely linked to local physical conditions in the wind, so it 
can absorb photospheric radiation and map out an 
expanding wind in a luminous star.

The lower level of the $\lambda$10830 multiplet ($1s 2s\ ^3S_1$) lies 19.7 eV 
above the ground state of \ion{He}{1} and can be populated directly by collisions
from the ground state ($1s^2\ ^1S$) although the rate is far less than 
for an allowed transition. High temperatures ($\ga$ 20,000 K) are generally required.
The $^3S$ level can also be populated by recombination from the continuum. This
latter pathway has long been studied especially in the Sun and
other cool stars (Goldberg 1939; Harvey \& Sheeley 1979;
Zirin 1975, 1982) because a source of EUV or X-ray radiation can 
photoionize \ion{He}{1} from the ground state or the $^3S$ level and
the helium ion  preferentially
recombines to the triplet state followed by subsequent cascade 
to the lower level of the $\lambda$10830 
transition. Thus stars with strong X-ray emission display  enhanced
helium absorption in $\lambda$10830 (O'Brien \& Lambert 1986; Zarro \& Zirin 1986;
Sanz-Forcada \& 
Dupree 2008).
Depopulation of the $^3S$ state occurs at high densities with collisions to the
$^1P$ level.  Because $^3S$ has a long lifetime for decay to the
ground state, 
(i.e. this level is metastable), a significant population can build up,
and provide an opportunity for scattering of near-IR photons from the line itself or
continuum photons from the photosphere. If the chromosphere is expanding, this transition
can trace out the  wind velocity as it scatters radiation while
being carried along in the expansion.  An additional advantage of this transition is that
the profile can not be compromised by interstellar or circumstellar absorption since
it is not a resonance line.

The first detection of a 
wind  in a metal-deficient
star using the $\lambda$10830 line was made  (Dupree et al. 1992a)  
in the bright field giant, HD 6833 where an
outflow of 90 km s$^{-1}$ was discovered  $-$ a  value 
comparable to the chromospheric escape velocity. 
Subsequently, Smith et al. (2004)
identified  He I absorption from one warm AGB star in the globular cluster M13
in addition to two other metal-poor field giants.  The short wavelength
extension of the lines in these stars reached 90 to 140 km s$^{-1}$ -- again fast
enough to escape a chromosphere and also a globular cluster.
A stellar $T_{eff}$ greater than 4600K appeared required to
populate the lower level of the He I atom; thus Smith et al. (2004)
suggested that   
the coolest red giants can not produce this 
transition.  Indeed, $\lambda$10830 was not detected in the 5 coolest red
giants observed in M13. While globular cluster stars themselves remain 
ideal targets, the metal-deficient field giants are brighter, more
accessible to current instrumentation, and can act as surrogates
for cluster stars.  We report here on the  
high-resolution spectroscopy of the \ion{He}{1} $\lambda$10830 line in 41 such
field stars.

\section{Observations and Reductions}

The objective in this investigation was to study the systematics of the \ion{He}{1} 
line among evolved Population II stars in a variety of evolutionary states.
This goal suggests an observational program concentrating on halo field stars
rather than globular cluster stars, since in the latter only the upper regions
of the red giant and asymptotic giant branches can be studied at high 
signal-to-noise, even with the NIRSPEC instrument on the Keck 2 telescope. Spectra 
of the \ion{He}{1} line for several red giants in the cluster M13 were published by
Smith et al. (2004). In our study, halo field stars were chosen from the lists of 
Bond (1980) and Beers \etal\ (2000) with an effort to achieve a good sampling 
in the red giant branch, red horizontal branch, and asymptotic giant branch 
phases of evolution. No selection was made on the basis of metallicity or
proper motion, although certain radial velocities were avoided in order to 
prevent overlap of the \ion{He}{1} line with telluric absorption features. Some
chemically peculiar stars in the form of CH stars were included in the sample.
To facilitate high signal-to-noise spectroscopy, only stars with apparent
magnitudes of $V < 11$ were observed, and most have $V < 10$. Thus this sample 
comprises relatively nearby halo stars, although most are beyond the limit at 
which the {\it Hipparcos} satellite provided reliable parallaxes. In order to 
study evolved stars, most of our targets have absolute magnitudes of
$M_V < +2$, and no subdwarfs were included in the program.

The spectra of 41 metal-poor halo field giants (Table 1) were obtained 
during 1.5  nights of observation in May 2005 using  NIRSPEC
(McLean et al. 1998, 2000)
on the Keck 2 telescope.  Observations were made using
the echelle cross-dispersed mode of NIRSPEC with the NIRSPEC-1 order-sorting
filter and a slit of $0.43^{\arcsec} \times 12^{\arcsec}$ giving 
a nominal resolving power of 23,600. The long-wavelength
blocking filter was not used in order to minimize unwanted fringing.
Total integration times accumulated for each star are listed in Table  2;
these times are generally broken into two shorter exposures in the NOD-2 positions.  
Calibration exposures consisted of 
internal flat-field lamps, NeArKr arcs, and dark frames.
Spectra of rapidly rotating hot stars obtained at airmasses
similar to the target objects were used to identify, and minimize 
or eliminate night sky emission lines.

Data reduction was performed using the REDSPEC package (McLean \etal\
2003) which was written specifically for NIRSPEC. Following dark 
subtraction and flat fielding,
the data frames were spatially rectified.  Wavelength calibration was
performed using NeArKr arc lamp spectra taken after each science
observation. For this study, we extracted only order 70 (wavelength
coverage $\sim 1.079 - 1.095$ $\mu$m), which contains the \ion{He}{1} line
at 1.0830 $\mu$m.  The strong \ion{Si}{1} photospheric absorption line
at 1.0827$\mu$m was used to estimate the stellar radial velocity with
an uncertainty of $\sim$5 km~s$^{-1}$.  Values of the radial velocity (RV)
are given in Table 2.

The spectrum of each giant
was normalized to a continuum determined by fitting
a 5th order cubic spline to the wavelength range 1.082 - 1.092 $\mu$m.
The wavelength scales of the spectra were then shifted onto the photospheric
rest frame of each star by applying a zero-point wavelength shift
determined from measuring the wavelength of the nearby photospheric 10827.09\AA\
\ion{Si}{1} line.  The equivalent width 
of the helium line
was measured from the continuum normalized spectra 
using the IRAF\footnote{Image Reduction and Analysis Facility (IRAF)
  written and supported by the IRAF programming group at NOAO,
  operated by AURA under a cooperative agreement with the NSF. 
({\it http://iraf.noao.edu/iraf/web})} 
task `{\it splot}' and measuring the width directly, or,
if blended with the adjacent Si I line, deconvolving the blend with
Voigt profiles and dividing the total equivalent width measured
directly into appropriate fractions obtained from the
deconvolution. The 
photometric colors, extinction, and evolutionary state as inferred
for the target stars
are included in Table 1. The footnotes to this Table contain the 
references to the tabulated quantities.  The assignment of the evolutionary state
of our targets is based on the Str\"omgren $c_1$ vs. $b-y$ diagram and
the color magnitude diagram ($M_V$ vs $B-V$).  Parameters of
the \ion{He}{1} line  are
given in Table 2.  About half (21) of the 41 targets are located on the
red giant branch (RGB), and one  (HD135148) is identified as a CH star.
Red horizontal branch (RHB) stars comprised 11 objects, and asymptotic
branch stars (AGB) made up 6 targets.  Two subgiants and one
semi-regular red 
variable (TY Vir) completed the sample.  The spectra of these stars are shown
in Figure  1 (RGB stars), Figure  2 (RHB stars),  Figure   3 (AGB stars), and
Figure  4 (subgiant stars). 

The stars in our sample show three basic types of 
He I line behavior: (i) a helium line with a pure absorption profile, 
(ii) a P Cygni type profile in which an absorption profile is paired with 
an emission feature to longer wavelengths, and (iii) no helium line at all - 
neither in absorption or emission.  Inspection shows that the  He I
line is generally broader than the neighboring Si I photospheric 
absorption line which
is expected since He I arises in the higher temperature  chromosphere.
In many cases where a He I absorption profile is present, this 
profile extends to shorter wavelengths providing evidence of a 
chromospheric outflow (see, for example, the spectra of stars
BD~$+$30\arcdeg2611 
and HD~122956 in Figure  1, BD~$-$03\arcdeg5215, HD 119516 in Figure  2, and
HD~121135 and HD~107752 in Figure  3).

The star HD 135148, classified as a CH star, and identified as  
a spectroscopic binary (Carney et al. 2003) exhibits (Figure  5) a substantial P Cygni
profile with deep absorption almost to zero flux at velocities $\sim-$60~\kms, 
and extending to $-$115 
~\kms.
The spectrum of the  coolest target of our sample, HD104207 (GK Com) is shown
in Figure  6 where many  photospheric absorption features of
neutral atoms appear in addition to a weak P Cygni feature of
He I.  The value of [Fe/H]=$-$1.93 in this star is similar to many
other
stars in our sample without such an array of neutral species in their
spectra, thus dramatically illustrating the effects of low effective temperature.

\section{Discussion}

The targeted  sample includes  $(B-V)$ colors ranging
from 0.6 to 1.6, and spans 6 magnitudes, reaching from the tip of the 
RGB to several subgiant stars.  Most of the stars  showed a \ion{He}{1} $\lambda$10830 
feature the presence of which is shown as a function of position
in the $M_V$ versus $(B-V)_0$ and T$_{eff}$  diagrams in Figure
7. The targets from M13 
and the metal-deficient field giants reported
earlier (Dupree et al. 1992a; Smith et al. 2004) have been added 
to this figure and their parameters are given in Table 3. 
Many of the more luminous stars  at $M_V=0$ and brighter, 
exhibit emission  resulting from scattering 
of the line above the stellar limb  which is normal  
in large stars with extended chromospheres. 
Since the \ion{He}{1} line is formed at chromospheric temperatures ($\sim$10,000$-$20,000K),
it might be expected to vanish in the coolest objects where the
chromosphere does not attain sufficiently high temperatures.  
A region exists in the color magnitude diagram,
with magnitude brighter than $-$1.5 and with  $(B-V)_0 > 1.1$  where 
the $\lambda$10830 line does not generally appear either in emission or
absorption.  Our earlier search for
\ion{He}{1} along the red giant branch in M13 revealed  absorption 
in an AGB star, IV-15, near $(B-V)_0$ =1.02, but not in any of  5 cooler 
stars on the red giant branch (Smith et al. 2004).  
The presence of a $\lambda$10830 line among Population II giants as
a function of position in a color-magnitude diagram 
differs from Population I stars (O'Brien \& Lambert 1986; Lambert 1987), 
where the helium emission disappears near spectral type M1 in 
giants and supergiant stars, corresponding to $T_{eff}$ of 3780 K
(Tokunaga 2000).  Based on our earlier sample (Smith et al. 2004), we 
noted that Population II giants with $T_{eff}$ less than 4600K did not show
helium.  However the larger sample presented here contains several stars with 
$T_{eff}$ less than 4600K, and they exhibit the $\lambda$10830 line. 
Two of these objects are somewhat anomalous red giants.  HD 104207 (GK
Com) is the coolest star in the sample, and a  semi-regular variable. It
is plausible that the atmosphere cycles through heating and cooling
phases, producing the helium line at certain times. The other 
star, HD 135138, is a CH object,   a spectroscopic
binary with a degenerate secondary star both of which could contribute to 
atmospheric   conditions of high excitation.  Yet a handful of otherwise 
normal stars with T$_{eff} <$ 4600K  remain: HD 6833; 
BD+30\arcdeg2611; HD 141531; and  HD 83212.  This extended survey of helium suggests
that only the most luminous Population II stars ($M_V$ brighter than $-$1.5), 
with $T_{eff} \lesssim 4500$ K lack the helium feature.

\subsection{Equivalent Widths}
The equivalent widths (EW) of the He I $\lambda$10830  absorption 
are shown in Figure  8 and 9 as a function of $T_{eff}$ 
and values are listed in Table 2. Repeated measurements suggest the
error in measuring the equivalent width is about 5\%.  The red giants may have an 
increasing equivalent width with decreasing effective temperature; this is
not unexpected as an extended expanding atmosphere increases 
scattering in the line. In the coolest star, HD 104207,  the He line
is blended with \ion{Ti}{1} absorption to shorter wavelengths.  Here, the
measurement of the equivalent width is uncertain because of the blend
with both \ion{Ti}{1} and \ion{Si}{1}.  A hint that the absorption is
extended and the equivalent width underestimated in HD 104207  comes from comparison
of the short wavelength side of another \ion{Si}{1} line at
$\lambda$10843.90 to that of the line at $\lambda$10827.09 (Figure  6). There may
be excess absorption on the latter line arising from an extended 
\ion{He}{1} profile.

The \ion{He}{1} line is surprisingly strong in the RHB stars.
  In 4 out of 11 RHB stars, the depth of the line extends  10 to
20 percent below the continuum and reaches equivalent widths between
0.1 and 0.5\AA. These  stars also show a dramatic
increase in the helium equivalent width that sets in at $T_{eff}
\gtrsim 5300$K.  The values of the  equivalent widths are comparable to those
found for  Population I stars including binaries which are
well-known X-ray sources (Zarro \& Zirin 1986; Sanz-Forcada \& 
Dupree 2008).  RHB stars are not known to be X-ray sources
(which would enhance the ionization of He I,  populate
the metastable level of \ion{He}{1} by recombination and cascade, and create 
a stronger line). 
There are no X-ray sources at the positions of the two stars
with strongest He absorption 
in the HEASARC Archives [http://heasarc.gsfc.nasa.gov] indicating that
X-ray illumination does not appear to be  present to strengthen the line. 
The RHB star, BD +17\arcdeg3248 has a chromosphere as documented also by
  the presence of \ion{Mg}{2} ultraviolet emission (Dupree \etal\ 2007).

The star HD 195636 displays an exceptionally  
strong He line. Preston (1997)  first noted
that this red horizontal branch star is a rapid rotator, which
is confirmed by Carney et al. (2008) as a single star. A strong helium 
line is present also in the
RHB object, HD 119516 which is not rapidly rotating and has
not been identified as a binary (Carney et al. 2008). 
Thus rotation does not appear related to the strength of the helium line,
although our sample consists only of  4 stars.  It is interesting to note that 
these horizontal branch stars are in a similar helium-burning evolutionary
phase as Population I clump giants. Clump stars, such as the
well-studied Hyades giants, exhibit magnetic activity cycles,
ultraviolet
emission, and X-rays (Baliunas \etal\ 1983, 1998).

The increasing strength of the \ion{He}{1} line with higher
effective temperature---possibly connected to the development of a
hotter chromosphere---suggests that collisions might be effective in
populating via the forbidden transition 
from $1s^2\ ^1S$ to $1s2s\ ^3S$. The strength
of the line in the CH star HD 135148, which has a degenerate companion,
demonstrates that ionization of He by a hot companion, with subsequent
recombination and cascade, can also be important in populating the
$1s2s\ ^3S$ level. Some fraction of stars will have undetected white
dwarf companions; this may be another parameter affecting the strength
of the helium line.

The absorption equivalent widths in this metal deficient sample 
are generally lower than found in bright Population I stars of low
magnetic activity. In Figure 10, the absorption equivalent widths for single red giants 
(Luminosity classes II-III, III, and III-IV) are taken from
the high quality measures in the sample of O'Brien and Lambert (1986).
Many of the spectra show variability in the equivalent widths, but
the \ion{He}{1} line is generally stronger in the stars with 
roughly solar composition than in the metal-poor sample. A P Cygni
profile may suffer some filling in of the absorption by near-star
scattering, but not all lines display these profiles. The helium
abundance, Y, increases by $\sim$20\% between metal-poor and solar
models (Girardi \etal\ 2000), which is not enough for a factor of
$\sim$4 change in the equivalent width.  For the same energy input,
the metal-poor chromospheres may be warmer since radiative losses are
less, however that would strengthen the helium line and not weaken
it. Perhaps the helium absorption is enhanced in the Population I
stars. They generally exhibit magnetic activity which leads to 
X-ray emission that increases the lower-level population 
through photoionization followed by recombination.  
Semi-empirical models of these chromospheres are needed.

The equivalent width of the $\lambda$10830 line 
as a function of metallicity is shown in Figure  11 where
no systematic dependence on [Fe/H] appears over this range of
lower [Fe/H]: $-$0.7 to $-$3.0.  Red giants of 
solar metal abundance show varying
strengths of the \ion{He}{1} line that tend to cluster between 100 and
200 m\AA, although several stars display values comparable to
the metal-deficient sample.\footnote{The stars that are 
X-ray sources among the Population I giants show substantially increased 
strength in the helium lines (Zarro \& Zirin
  1986; O'Brien \& Lambert 1986; Sanz-Forcada \& Dupree 2008),
  presumably due to X-ray photoionization, followed by recombination
contributing to the population of the lower $^3S$ level.}
 It is premature to speculate on the 
helium abundance from the line strengths alone without modeling this 
and other helium profiles,
since they depend on chromospheric conditions.  However, if spectra
could be obtained in a globular cluster, providing a larger sample of
similar stars, the relative abundance of helium might be assessed.

\subsection{Line Profiles}

About one-third of all the luminous stars in Figure  7 (AGB and RGB stars
brighter than $M_V = 0.5$)  show
helium emission. In most of these stars,
the emission is accompanied by absorption. These  classical P Cygni
profiles by their very nature mark an extended outflowing atmosphere.
Absorption profiles without emission can indicate atmospheric dynamics
by their asymmetry.  The ratio of the short wavelength extent (at the
continuum
level) to the long wavelength extent of an absorption feature relative
to the photospheric rest wavelength gives a measure
of the line profile asymmetry. These values are converted to velocity
units and are shown in Figure  12 for the
stars without emission.  The value of  $B/R$ is given where $B$
denotes the blue (short wavelength) extent and $R$ the red (long wavelength) extent.  
The majority of the helium lines have $B/R>1$ signaling outflowing motions. 
Values of the
short wavelength extent of the \ion{He}{1} absorption are taken as
the terminal velocity ($V_{term}$), and $B/R$ ratios are given in Table 2.
It is generally easy to see from the spectra that 
helium absorption can extend to the strong \ion{Si}{1} line at 10827.09\AA.  
Such an extension implies an expansion velocity of at least 
90 km s$^{-1}$.  Many stars exhibit higher speeds with absorption
evident in the short wavelength wing of \ion{Si}{1}.
The outflow velocities measured by the extent of the short wavelength
wing are independent of [Fe/H] as shown in Figure  13. Metallicity,
naturally, is a factor determining the speeds of radiatively-driven winds. And there is
some evidence for dusty winds in OH/IR sources in the 
low metallicity Magellanic clouds to have lower speeds 
as compared to similar sources in the galactic center
(Marshall \etal\ 2004). However, the outflow speeds of gaseous winds 
detected here (presumably driven by hydrodynamic or magnetic
processes) do not depend on the [Fe/H] abundance, and we conclude that
these winds are not radiatively driven.

RHB stars have a convective core and a semi-convective envelope
(Castellani et al. 1971; Schwarzschild 1970) and so conditions exist 
for the acceleration of a stellar wind by 
magnetic processes such as Alfv\`en waves.  In addition, 
if high temperatures are produced  in an extended chromosphere, these  could contribute
to a thermally driven wind.

The extension of the 10830\AA\ line to shorter wavelengths signals
outflow that in many stars is comparable in value to the escape 
velocity from the stellar chromosphere: 

\begin{equation}
  V_{esc}(\rm km\ \rm s^{-1}) = 620 \left( \frac{{\it
  M/M_{\sun}}}{{\it R/R_{\sun}}} \right)^{1/2},
\end{equation}

\noindent
where $M$  is the stellar mass and $R$ is the distance from the star
center to some region in the chromosphere. 
We take the 10830\AA\ line to be formed at 2R$_\star$ in the stellar
chromosphere where R$_\star$ is the stellar photospheric radius.
This estimate does not require a helium model.  Detailed calculations 
concur on the location of the formation of H$\alpha$ in metal-poor stars.  
Observations of higher outflow velocities as well as semi-empirical models confirm that the 
10830\AA\ line is formed above the H$\alpha$ core in luminous stars.  
Our spherical models for metal-deficient  giants (Dupree et al. 1984) have
a chromospheric extent of 'several' stellar radii in order
to produce the H$\alpha$ line (1.2 R$_\star$ to 3.6R$_\star$). 
The  recent (non-LTE, spherical expanding) models of 
Mauas \etal\ (2006) note that the H$\alpha$ core is formed 'about 1 
stellar radius above the photosphere', similar to our spherical 
models. Subsequent modeling of H$\alpha$ in M13, M15, M92 giants show 
the H$\alpha$ cores to be formed at ~2R$_\star$ (M\'esz\'aros et al. 2009a).
Hence, it appears reasonable to assume the level of formation 
of the 10830\AA\ line as 2R$_\star$.  Table 2
contains an estimated stellar radius (Column 10)  determined by
evaluating the
bolometric correction for each star (Alonso \etal\ 1999) 
as a function of $T_{eff}$ and [Fe/H].  The chromospheric escape velocity
for each star is tabulated in Column 11 of Table 2 using 
Equation (1). Here we have
assumed masses of a red giant (0.75M$_\odot$), a red horizontal branch
star (0.7M$_\odot$), an AGB star (0.6M$_\odot$), a subgiant branch
star (0.8M$_\odot$), and a semi-regular variable (0.6M$_\odot$). 
Marked in boldface are  values of $V_{term}$ when 
they are comparable to or exceed the escape
velocity at 2R$_\star$.  These amount to 40\% of our sample where
the helium line is detected.

Many of the remaining stars exhibit a short wavelength extension 
of $\gtrsim$40 \kms; this value exceeds the extension of the long
wavelength wing, and signals that outflow of material is present that
has not yet reached escape velocity.   Where helium occurs in the luminous
stars, ($M_V \lesssim -0.2$), a signature of outflow is found in each one.  
However, a fraction of the lower luminosity objects do not show this 
signature, suggesting that the gas outflow may be variable.\footnote{One star, 
BD+17\arcdeg3248 ($M_V=0.65$), was observed 3 years previously (Smith et al. 2004) and
the helium profile has not changed.}  Speeds greater than $\sim$
12~\kms\   are generally supersonic in fully ionized metal-deficient
([Fe/H]=$-$2) plasma at chromospheric temperatures of 10$^4$K.

Three of the target stars (BD +17\arcdeg3248, HD 122956, and HD
126587)
have high resolution Hubble Space Telescope spectra available of the \ion{Mg}{2} 
line at $\lambda$2800 (Dupree et al. 2007). Asymmetries of the line 
emissions indicate motions in the chromosphere, and these were found
in stars brighter than M$_V$=$-$0.8.  Only the most 
luminous of the 3, HD~122956, shows a
\ion{Mg}{2} emission asymmetry indicating outflow (short-wavelength 
emission peak less than
the long-wavelength emission peak).  HD 122956 has a high value of the
helium terminal velocity, 110 \kms\ which exceeds the chromospheric
escape velocity of 78 \kms.   The RHB star, BD +17\arcdeg3248, shows outflow in
helium at an intermediate velocity, whereas the helium line in HD 126587
appears symmetric. However, models (Dupree et al. 1992a) 
suggest that \ion{Mg}{2} is formed at lower levels than the 10830\AA\
line in a metal-deficient chromosphere.\footnote{Other cool stars have
observational signatures of this separation: Cepheids (Sasselov \&
Lester 1994), a T Tauri star (Dupree et al. 2005), and 
the Sun (Avrett 1992).} Thus it is not surprising to
find differing dynamical signatures in these line diagnostics, in
addition to possible time variations.   There
may be a similarity here between the well-known changing asymmetries of  H$\alpha$
emission wings in metal deficient giants 
(Smith \& Dupree 1988; Cacciari et al. 2004; M\'esz\'aros et al. 2008,
2009b) and the \ion{Mg}{2} emission.

The star HD 135148 deserves special mention. This RGB object 
is classified as a CH
star and Carney et al. (2003) obtained an orbital period of 1411 days
for the spectroscopic binary.
Emission in the P Cygni profile arises from an extended scattering
atmosphere and the absorption extends to $\sim$$-$115 km s$^{-1}$. This
value exceeds the escape velocity from the chromosphere,
$V_{esc}$=67 \kms, where helium originates.  Thus
after the initial transfer of material to the secondary star in the
system,  a substantial wind remains from the cool star that is presently visible.

\subsection{Estimate of the  Mass Loss Rate}
A rough estimate of the mass loss rate implied by 
the helium absorption in the wind can be derived from the Sobolev optical
depth.  In an expanding atmosphere, a photon emitted from the
photosphere (or the line itself) can be absorbed and then scattered 
when the absorption coefficient is  ``aligned'' with the photon. In 
this situation,  a sufficient number of atoms occur at the
correct velocity to  absorb and scatter  photons.  In
a stellar wind, a narrow interaction region will be present that 
depends on the velocity gradient in the wind, and the width and
strength of the line absorption coefficient.  The Sobolev 
approximation defines the interaction region to be very narrow  
for simplification of the transfer equation, and this causes the
absorption parameters  to be related only to local conditions
(Lamers \& Cassinelli 1999).
This approximation assumes that the  density and velocity
gradient  do not change significantly 
over the absorbing/scattering region.

The line optical depth at frequency, $\nu$, at star center, is given by
\begin{equation}
\tau_\nu = \int_0^\infty \kappa_\nu(z) \rho(z) dz
\end{equation}
along the radial direction, $z$, where $\kappa_\nu$(cm$^2$~g$^{-1}$)
is the 
line absorption coefficient, and $\rho$(g~cm$^{-3}$) is the mass density
in the lower level of the transition.  Taking the line profile
function
as a delta-function (the Sobolev approximation) and inserting values for $\kappa_\nu$ 
[cf. Equation (8.51) of Lamers \& Cassinelli (1999) or Equation (8.8) of
  Hartmann (1998)], we write 
the Sobolev optical depth, $\tau_S$ as:
\begin{equation}
\tau_{S} = \frac{\pi e^2}{m c} \times f \times \lambda_0 \times \frac{N_1}{(dV/dz)}.
\end{equation}
where $dV/dz$ is the velocity gradient in the scattering region and 
$N_1$ is the population in the lower level of the absorption
line (the $^3S$ level of \ion{He}{1}).  Conservation of mass gives 
\begin{equation}
\dot M = 4 \pi R^2 V \mu m_H N_H(V) 
\end{equation}

\noindent
where $R$ is the radial distance (in units of $R_\sun$) at which the wind has a
velocity $V$(km s$^{-1}$).  N$_H$ is the hydrogen density at $V$, $\mu$
is the mass per hydrogen nucleus, and m$_H$ is the mass of the hydrogen atom.
Then substituting for $N_H$ into  Equation (4) 
[rewriting $N_H = N_1 \times (N_H/N_1)$], and 
replacing the value of  $N_1$ from the expression for $\tau_S$ above, we find

\begin{equation}
\dot M = \frac{4\pi R^2 V \mu m_H}{N_1/N_H} \times
\frac{\tau_S}{\frac{\pi e^2}{mc} f \lambda_0} \times \frac{dV}{dz}.
\end{equation}

\noindent
We assume that $ dV/dz \sim \Delta V/\Delta R = V/(R-R_\star)$ where $\Delta V$ 
is the change in wind velocity,  ($R-R_\star$) is the distance over which the speed changes 
from zero at the stellar photosphere to a value of $V$ at distance $R$. In our estimate, we adopt 
$R=2R_\star$ since the 10830\AA\ line is formed in the chromosphere
(see discussion in Section 4.2) and $R$ is measured from the
center of the star.  Then 
with  $\mu=1.4$, and $m_H=1.67\times 10^{-24}$ g, 
we have:

\begin{equation}
\dot M\ {(\rm M_\sun\ \rm yr^{-1})} = \frac{1.22 \times 10^{-18} \tau_S (R/R_\sun)^2
  V^2}{(N_1/N_H) \times f \times \lambda(\AA) \times (R_\star/R_\sun)} 
\end{equation}

\noindent
where $N_1/N_H$ is the
ratio of the population in the
lower $^3S$ level of the helium 
transition to the total hydrogen density. 
To evaluate the mass loss rate from Equation (6)  at a
distance of 1 $R_{\star}$ above the photosphere,
we set $\tau_S$~=1 and the oscillator strength, $f=0.54$
for the $\lambda$10830 multiplet.  With
these values, the mass loss rate becomes,
 
\begin{equation}
\dot M\ (\rm M_{\sun}\ \rm yr^{-1}) = \frac{8.37 \times 10^{-22}
  {\it R_\star V^2_{term}}}{{\it (N_1/N_H)}} 
\end{equation}

\noindent 
where $R_\star$ is the
stellar (photospheric) radius (in units of $R_\sun$), and $V_{term}$
is the observed terminal velocity (km s$^{-1}$) in the helium line.

Now, the value of $N_1/N_H$  
can be estimated  
using our semi-empirical non-LTE  models
(Avrett \& Loeser 2008) of cool star
chromospheres  (see Appendix A).  The semi-empirical models suggest
that an upper limit of the population ratio, $N_1/N_H$ for typical line strengths 
found in the targets reported here (line depth $\sim$0.9) corresponds
to 6.3$\times$10$^{-8}$ where $N_{He}/N_H = 0.1$.
As a lower limit, we take the value derived from the solar model:
1.0$\times$10$^{-8}$.  The mass loss rates are estimated using values
for $V_{term}$ (where $V_{term} > 45$\kms) and $R_\star$ contained in
Table 2, and are shown in
Fig. 14.  Two rates are plotted for each star corresponding to the
upper and lower limit on $N_1/N_H$. The mass loss rate generally
increases with increasing stellar bolometric magnitude.  
The uncertainty in these estimates arises predominantly from the population
of the lower $^3S$ level of the $\lambda$10830 transition. A discussion of the
error estimate is given in Appendix B.  Equation (7) does not apply to
the exceptionally deep P~Cygni profile of HD~135148 such as that shown in Fig. 5
because it likely overestimates the mass loss rate.

Undoubtedly there is variation in the gas mass loss rate for  stars on the
RGB  - and probably all stars considered here.  
The solar mass loss rate changes over the solar cycle by
about a factor of 1.5 (Wang 1998).  M\'esz\'aros \etal\ (2009a) found 
variations ranging between a factor of 2 to 6 in the mass loss rate of 
individual globular cluster red giants.  Accompanying the changes 
in the mass loss rate may be  variations in the size of the outflow
velocities. However several observations suggest that outflows occur 
continually as stars evolve through the upper part of the RGB.
The helium line profiles reported here 
overwhelmingly display a signature of outflowing gas 
in stars brighter than $M_V \sim 0$. 
H$\alpha$ line cores show generally increasing outflowing
velocities with luminosity (M\'esz\'aros \etal\ 2008, 2009b). 
Even though some measured outflow velocities are less than the
escape velocity, conservation of mass suggests that the velocities 
will yield meaningful mass loss rates.  We emphasize the difference
between the presence of mass loss from red giants 
determined from diagnostics of the gas (from  \ion{He}{1} 10830\AA\ and
the H$\alpha$ line) and that derived from infrared detections of 
circumstellar dust.  Evidence suggests 
that dust formation is an episodic process (Origlia \etal\ 2007;
M\'esz\'aros 2008), whereas the velocity measurements of the
gas indicate continuous outflow of material.

With regard to the presence of a wind directly indicated by the 
helium line profile, a fraction of the stars in Table 2 and 3 
show velocities exceeding the chromospheric escape velocity.  
There are several possible interpretations of these measurements: 
(1) Only some fraction of the stars have outflows
that develop into winds; (2) All stars develop winds for some 
fraction of their evolutionary phase brighter than $M_V \sim 0$.  
(3) The outflow velocities vary, probably along with the mass 
loss rates, and we need diagnostics formed
higher in the atmospheres to attempt the detection of escape velocities.  The
absence of inflowing velocities both in H-$\alpha$ and the helium line
would suggest that the last option appears to be the  most likely. 

\subsection{Evolutionary Effects of Mass Loss}

Stars on the red giant branch exhibit mass loss rates that increase
with luminosity.  At a magnitude comparable to the horizontal branch,
$\dot M \sim 3.2\times 10^{-9}$ to $3.2 \times 10^{-8}$ \smpy.   At 
higher luminosities on the RGB, the mass loss rates increase by about
a factor of 2, viz., $6.3 \times 10^{-9}$ to $6.3 \times
10^{-8}$\smpy. A low-mass star spends about 50 Myr in the RGB 
phase, thus, taking the geometric mean of the minimum and maximum
values, 
($4.5\times 10^{-9}$\smpy),   the total mass lost as indicated by the helium line 
would amount to $\sim$ 0.2 $M_\sun$.  This is the amount generally
demanded by stellar evolution considerations which range from
$\sim$0.15 to 0.22$M_\odot$ (Rood 1973; Lee \etal\ 1994; Caloi \&
D'Antona 2008; Dotter 2008). Moreover, this mass loss rate is
consistent with values derived from the H$\alpha$ profiles of globular cluster
giants (Mauas \etal\ 2006; M\'esz\'aros \etal\ 2009a). 

Stars on the AGB have estimated 
mass loss rates of 0.25$-$1.6$\times$ 10$^{-8}$ \smpy. 
For a 20 Myr lifetime on the AGB, additional
mass loss of $\sim$ 0.1M$_\sun$ could result for the AGB objects here 
(assuming a mean mass loss of 6.3 $\times$ 10$^{-9}$\smpy).
Groenewegen \& deJong (1993) find about 0.16$-$0.2 $M_\odot$ is lost on
the AGB.  The sample of AGB stars included here has low luminosities
($M_{bol} \approx$ 0 to $-$1.9), and it is likely that the mass loss 
increases with luminosity  which would increase the empirical total
mass loss.

Minimum and maximum values of the mass loss rate for the RHB stars 
in our sample range from $4.5 \times 10^{-10}$ to $2.2 \times 10^{-8}$
\smpy (Fig. 14).  Taking a median of $3\times 10^{-9}$\smpy, and 
a lifetime of 75 Myr, this rate implies a total mass loss of
0.2$M_\odot$.
Note that only half of the sample of RHB stars has an asymmetric
profile indicating outflow velocities, suggesting that the outflow
and hence the mass loss rate may be lower.  Some evolutionary models 
have considered mass loss on the RHB. Yong et al. (2000) and earlier,
Demarque \& Eder (1985) 
evaluated models of horizontal branch stars with the {\it ad hoc}
assumption of mass loss 
to test its  effects, 
and concluded that  mass loss rates between 10$^{-10}$
and 10$^{-9}$ \smpy\ could produce 
the observed extended blue HB.  Koopmann et al. (1994) set an upper limit
to the HB mass loss rate of 10$^{-9}$ \smpy, based on 
calculations for M4.  These values are not substantially discrepant 
from the mass loss rate inferred from the \ion{He}{1}  line profiles.

Vink and Cassisi (2002) evaluated the effect of
radiatively driven winds in horizontal branch stars, but found the mass loss
rates too low for evolutionary effects.  However, winds can be
driven in several other ways which is likely here, given the presence
of convection layers in these stars (Castellani \etal\ 1971;
Schwarzschild 1970).  It is worth noting that even  
models of RGB and AGB stars currently consider sub-surface dynamo activity
(Busso et al. 2007, Nordhaus et al. 2008) which  could lead
to winds driven by magnetic processes. 

Carney \etal\ (2008) noted that 
red horizontal branch stars in the field are rotating slower than
expected 
considering the rotation observed in their predecessors on the red 
giant branch.  Angular momentum carried away by the stellar winds
could offer an explanation of  this discrepancy. Here we estimate
the torque required to spin down a red giant with the mass loss rates
discussed above. To be effective, magnetic fields must be present and we assume that
the moment of inertia of the  star does not change. The torque
required, $\tau_\star$ (dyne cm)  under these conditions is
$\tau_\star=(I_o\omega_o)/\Delta t$ where $I_o$ is the stellar moment of
inertia,  $\omega_o$, the initial angular velocity, and $\Delta t$ the time
required to completely stop the rotation of the red giant. Taking,
a tangential rotation velocity of 2 km s$^{-1}$ [from the Carney \etal\ (2008)
  measures, and setting $sin\ i =0.3$]  and R$_\star$=50R$_\sun$ for
a 0.8M$_\sun$ red giant star, we find that a 
torque of 7$\times$10$^{35}$ dyne cm is required 
to spin down the star in 20 Myr.  Matt \& Pudritz (2008) evaluated
the torque created on a star by  winds of  various
mass loss rates in the presence of  a stellar magnetic field with
several configurations and strengths.  A simple parameterization
of the quantity: $(B_\star R_\star)^2/(\dot M V_{esc})$
can predict the 'lever arm' to evaluate the torque
exerted by the wind on the star.  For a red giant with the
parameters above,  a dipole magnetic field of 100 gauss, and  $\dot  M$= 4.5
$\times$10$^{-9}$ \smpy, the predicted wind torque amounts 
to 3 $\times$ 10$^{36}$ dyne
cm, a value that exceeds the amount required for spindown, making winds a
plausible explanation of the low velocities of RHB stars.  However
we  caution that a change in the moment of inertia could
affect  these results as well as the magnetic field strength and
configuration.\footnote{Y. C. Kim made available his calculations of
  the moment of inertia for a 1~M$_\sun$
  star with [Fe/H]=$-$0.25 as it evolves up the red giant branch.  These
  show that the moment of inertia increases by a factor of 6.5, based on internal
structural changes including an increase in radius,  as the
star evolves from an effective temperature of 4500K to 4000K.  Such a
factor would  appear to allow spin-down to occur.}

These results can be used to calculate a total budget of mass loss for
post-main sequence evolution. Taking values of $\dot M$ estimated above,
we find a star would lose 0.2~M$_\sun$ on the RGB, 0.1--0.2~M$_\sun$ on the
RHB, and 0.1--0.2~M$_\sun$ on the AGB and as a planetary nebulae (Bianchi
\etal\ 1995), totaling 0.4--0.6~M$_\sun$. This total is in harmony with
recent studies of globular clusters that suggest anywhere from 0.5 to over
1~$M_\odot$ is lost between the main sequence turnoff and the white dwarf
cooling sequence (Moehler \etal\ 2004; Hansen \etal\ 2007; Richer \etal\
2008). Our budget should not be strictly taken to apply to current turnoff
stars of mass $\sim 0.8$ M$_\sun$, since many of the white dwarfs observed
in clusters are remnants of stars of higher initial masses, where more
mass loss is demanded in post-main sequence evolution. In addition, the Population II
field giants that make up our sample potentially had a range of initial
masses, and thus some may have mass loss rates higher than expected for
globular clusters.

\subsection{The Fate of Wind Material}

If red giants in globular clusters have winds similar to those 
detected among Population~II field giants, it is worth considering whether 
the wind material would have enough energy to escape not only from the stars 
themselves but also from the parent cluster.
Escape velocities from the cores of Milky Way globular clusters have been 
evaluated by McLaughlin and van der Marel (2005).  Central  escape
velocities vary from $\sim$ 2 \kms\ (AM1, Pal 5 and Pal 14) to $\sim$
90 \kms\ for massive clusters (NGC 6388 and NGC 6441).  A majority of the stars
brighter than M$_V$ = 0 in our sample exhibit terminal velocities in excess of
the stellar escape velocity from the chromosphere. These 
speeds, ranging from 90 to 170 \kms\  also exceed many of the escape speeds from clusters too. 
Since the cluster escape speed corresponds to the speed necessary
to remove material from the core,  the
value will obviously decrease with distance from the core. 
One could envision a scenario in which material
does not have sufficient energy to escape the cluster core,
but it could escape if arising from a star located further out in the cluster.

Several authors have considered in detail the dynamics of wind
material deposited in the intracluster medium by the stellar 
population.  VandenBerg and Faulkner (1977) used hydrodynamic
equations to construct time
dependent models of gas flow from a cluster assuming the stellar wind
has a velocity $\sim$20 \kms.  Depending upon the
initial 
assumptions of the energy available, they found that both  
outflows and inflows of material could occur, the latter as a result
of radiative cooling. In some models material
was retained in the cluster core. Smith (1999) concluded that additional energy could be
injected into the gas by solar-like winds originating from cluster 
dwarf stars and this would be sufficient to establish an outflow of 
material from a cluster even if red giant winds are slow. 
If the giant winds are fast, comparable to the solar wind with V~$\sim$~450~\kms, 
steady state outflows would result, even from clusters 
with the highest values of escape velocity (Faulkner \& Freeman 1977).

The results presented here from helium lines suggest that the winds
from red giants and red horizontal branch stars can be
much faster than generally  assumed, and these would naturally lead to
a steady-state outflow.

Recent observations of the fast solar wind suggest that interactions between
colliding winds within the cluster itself may also be of some consequence.
Insight can be drawn from the Sun where
spacecraft have located the termination shock  and  characterized its
physical parameters (Richardson et al. 2008).  The shock occurs 
$\sim$100 AU distant from the Sun, and the wind speed at that point is 
comparable to the
speed in the corona, namely 400 \kms.  While the solar wind is hot and
driven by a combination of gas and wave pressure, a cool wind can be driven
by wave pressure alone (Cranmer 2008).  However, convective envelopes
in red horizontal branch stars, and recent conjectures of dynamo
activity  in RGB and AGB stars would allow MHD processes to occur as
well (Busso \etal\
2007; Nordhaus \etal\ 2008). Thus, it appears possible that the 
stellar winds will not be substantially decelerated up to the 
termination shock, but maintain the  values indicated by the helium line.

At large distances from the star, the
pressure of the stellar wind will eventually be
balanced by the opposing pressure (both gas and magnetic) presented by
the surrounding intercluster medium, $P_{icm}$.  When these pressures are equal, a
termination shock occurs at some distance, $R_{TS}$, viz.:  

\begin{equation}
R^2_{TS} = \frac{\dot M_\star \times V_{wind}}{4 \pi P_{icm}}.
\end{equation}
Rewriting this using astronomical units,
\begin{equation}
R_{TS}(AU) = 48.4 \times \sqrt{\frac{\dot M_\star(M_\sun\ yr^{-1}) \times
    V_{wind}(km\ s^{-1})}{P_{icm} (dyne\ cm^{-2})}}
\end{equation}
and assuming $\dot M=$2$\times$10$^{-9}$ M$_\sun$\ yr$^{-1}$ for an
average RGB star, V$_{wind}$=
100 \kms, and a (total) interstellar  pressure of 4.0$\times$10$^{-13}$
dyne~cm$^{-2}$ for a distance of 1.5 kpc above the Galactic plane (Cox
2005), we
find the transition shock distance is 34,000 AU or 0.2 pc from the
star. The central luminosity densities for globular clusters
have  median values of $\sim$~3000~$L_\sun pc^{-3}$ (Harris 1996), 
and with a  typical star of
0.1 to 0.2 $L_\sun$, the median stellar density in the core is greater than 10$^4$
stars pc$^{-3}$, giving an average separation between stars of less than
9500 AU - smaller than the termination shock distance.   While
uncertainty exists in
both the pressure outside and within 
clusters\footnote{The pressure in the galactic halo is not firmly known.
O VI absorption suggests warm (3 $\times$10$^5$K) extended low density
regions are present at high galactic latitudes 
(Dixon and Sankrit 2008) with thermal
pressures of 0.7 to 1 $\times$ 10$^{-12}$ dyne cm$^{-2}$.
Electron densities in the intercluster material of 47 Tuc 
derived from pulsar dispersion measures
(Freire et al. 2001) indicate
$n_e=0.067\pm 0.015\ cm^{-3}$. They concluded ionized material was
dominant, where for a temperature of 10$^4$K, the thermal
pressure equals 9$\times$10$^{-14}$ dyne\ cm$^{-2}$. van Loon et al (2006) detected
0.3$M_\sun$ of material in the core of  M15 using
the Arecibo telescope.  Assuming that this material is 
evenly distributed in the beam 
we find a hydrogen density of 2.29$\times$10$^{-2}$ cm$^{-3}$ 
for the diffuse gas in the  core. With a  temperature of 
100K, the pressure will be 3.2$\times$10$^{-16}$ dynes cm$^{-2}$.
Faulkner and Freeman (1977) constructed time-independent gas
flow models for tightly bound clusters.  These models suggest
pressures at the tidal radius ranging from 0.1 to 5.5 $\times$10$^{-16}$
dyne cm$^{-2}$. At the sonic point in the flow which lies in the cluster interior  where the
stellar density is down by a factor of $\sim$100 from the core 
density, the gas pressures are much higher, ranging from
5$\times$10$^{-15}$ to 3.5$\times$10$^{-12}$ dynes cm$^{-2}$.
Our estimate of the
transition shock distance varies inversely as the
square root of the external gas pressure, so that an order
of magnitude change in the gas pressure, changes the distance
by a factor $\sim$ 3.},
these estimates suggest that the central regions of clusters
may well be filled with expanding warm material, enveloping the surrounding stars. Since
the majority
of cluster stars are still on the main sequence  with
winds
of $\dot M \sim 2 \times 10^{-14}$ \smpy\  and $V_{wind} \sim
400$ \kms\ (adopting solar parameters), these winds can not balance
the pressure of the more massive winds arising from the luminous
stars, raising the
possibility that the dwarf winds are in fact smothered by the giant
winds, allowing for pollution of the surface layers of the
dwarfs. 
As the stellar density decreases towards
the cluster edges, the effect of smothered winds and resultant surface
pollution would decrease, possibly leading to spatially dependent
self-pollution within a cluster. Even beyond  the termination shock
crossing, 
the wind speed decreases by
about a factor of 2 (based on the solar termination shock measured
by Richardson et al. 2008), but the temperature and density increase.
In many cases, even the decreased speed  will allow escape of these
fast winds from the cluster.

Central escape velocities for the most massive clusters 
can reach $\sim$90 \kms\ (McLaughlin \& van der Marel 2005).   
While the helium line profile suggests that some objects
possess these high  velocities, others do not (see Figure 13).  The
extent
of velocity variability in the helium line  is unknown at present.  
The velocity of a red 
giant wind at levels higher than the formation region of the  helium 
line may (or may not) reach escape velocities.  
Calculations of line forming regions for strong optical, IR, 
and near UV lines have
been made for  metal deficient red giants (Dupree et al. 1992a; Mauas
et al. 2006; M\'esz\'aros \etal\ 2009a) but only through the low chromosphere and  
not for higher levels of the atmosphere.  In the high chromosphere, 
the most straightforward diagnostics of winds lie in the ultraviolet
region of the spectrum, where Lyman-$\alpha$  and resonance lines of C II
($\lambda$1335) and Si II ($\lambda$1800) might be 
observable.\footnote{UV and Far-uv spectra obtained 
with {\it IUE}, {\it HST}, and {\it FUSE}
of the brightest of the metal-deficient
targets in the field, HD 6833, are weak and can not definitively reveal
emission.}  One advantage of
metal-poor targets is that their high velocities move the spectra
away from local interstellar absorption which could compromise the resonance
line profiles.
Infrared transitions of hydrogen  such as members of the
Paschen, Brackett, and Pfund series arise from higher levels of
hydrogen and hence are formed  deeper in the atmosphere  and can not
help in detecting a wind.

\section{Conclusions}

The He I 10830\AA\ transition maps atmospheric dynamics to higher
levels than optical or near-uv diagnostics.
The profile of this near-infrared line in metal-poor stars gives 
evidence for outflow in most of the RGB, AGB, and RHB stars showing helium.
In many objects, the speeds are comparable to the escape velocities from both
the stellar chromosphere  and  globular clusters. The fact that all the 
luminous stars with helium absorption exhibit expanding chromospheres 
suggests that the mass outflows as traced by gas occurs
continually, with most probably variable mass loss rates. 

Our estimate suggests that the mass loss observed directly 
on the RGB will provide the requisite amount needed 
by stellar evolution calculations.  Mass loss detected in RHB stars 
appears at a rate sufficient to cause extension  of the
horizontal branch. It will be very useful to obtain ultraviolet 
spectra of some of these stars
to identify higher temperature plasma, and to track the
acceleration in the chromosphere.
Better estimates of mass loss rates 
can be obtained with semi-empirical modeling of the line profiles.

These results demonstrate that chromospheric material has sufficient speed to
escape these stars and become a stellar wind. 
If the star were in a globular cluster, the 
high-speed wind continues with little diminution, filling the cluster
with expanding warm gas. In the process, red giant winds
could smother the substantially less massive  winds from dwarf stars
possibly allowing for surface pollution.  Beyond the termination shock,
velocities decrease but could still escape the cluster.  
Thus fast winds observed in the helium
line offer a straightforward way to
understand the absence of  intracluster material in globular clusters.

\acknowledgments
We are grateful to Steve Cranmer and Aad van Ballegooijen for insight
into the solar wind. Gene Avrett kindly made the helium calculations
for the Sun available before publication. We thank Y. C. Kim for providing us with
his pre-publication results for the evolution of the moment of inertia
of red giants.    This research has made use of NASA's
Astrophysics Data System Abstract Service and SIMBAD database (CDS, Strasbourg,
France).  We wish to extend special thanks to those of
Hawaiian ancestry from whose
sacred mountain of Mauna Kea we are privileged to conduct
observations.  Without their generous hospitality, the Keck results
presented
in this paper would not have been possible. 

{\it Facilities:} \facility{Keck 2 (NIRSPEC)}

\newpage

\appendix

\section{Estimation of Level Population}

The level population for the $^3S$ level of \ion{He}{1}, n($^3S$), is estimated from
non-LTE calculations using semi-empirical models 
for different stars.  A detailed discussion
of helium excitation processes in the solar atmosphere is given
in Andretta \& Jones (1997) and is not reviewed here.  Our estimates
of the population levels derive from  the PANDORA code 
(Avrett \& Loeser 2003) which was used
in either plane-parallel or spherical form, with an expanding wind 
in most models. A 13-level
helium atom plus continuum was generally used; some models
had a 5-level helium atom plus continuum to expedite
calculation.  The velocity field was introduced explicitly into the source functions.
Solutions for hydrogen populations and ionization 
are iterated first, and then they are followed by similar iterations
for helium.   The highest value n($^3S$)/n($He_{tot}$) in each model is shown
in Fig. 15 as a function of the maximum depth of  absorption in the
helium 10830\AA\ line. The solar model represents the
quiet sun for a static plane-parallel
atmosphere which is described in detail elsewhere (Avrett \& Loeser 2008),
and E. Avrett kindly made an advance copy of the helium results  
available for use here.  High
It may well be that the mass loss rates are variable. For instance only 
one-half of the RHB stars exhibit outflow. And recent chromospheric models
calculated for varying H$\alpha$ profiles in globular cluster red
giants, reveal changes in mass loss rate by a factor of 6 (Meszaros
\etal\ 2009a).  Thus the total mass lost would seem to intrinsically span a range
of values.  
luminosity stars (similar to giants and supergiants) were also calculated
in plane-parallel and  spherical models both in a  static atmosphere and 
also with an assumed mass outflow. The models were of cool
stars (such as $\beta$ Dra, $\alpha$ Aqr, and $\alpha$ Boo)
with effective temperatures approximately solar.  The photospheric
model has little or no influence on the helium line profile since it 
is formed totally in the chromosphere.   Results for Mg II and helium 
lines in a supergiant were described elsewhere (Dupree et al. 1992b).
A cool dwarf star with an extended chromosphere (TW Hya) producing a
P Cygni profile was also 
modeled in helium, with  a spherical approximation and a substantial
high velocity wind (Dupree et al. 2008).

Figure 15  shows the maximum  depth of the \ion{He}{1} 10830\AA\ as a function
of the population ratio n($^3S$)/n($He_{tot}$) of helium. The
three plane parallel models (for the Sun and the supergiant and giant
stars marked with vertical lines in Figure 15) show generally lower
values of the population ratio.  As one might
expect,  there is a
correlation between the depth of the line and the level population. 
Higher level populations result in increasing the line depth. The
models with high outflow velocities in the chromosphere (110$-$200 km s$^{-1}$)
at the atmospheric level where the $^3S$  population  maximizes, 
create deeper absorption profiles, but the level populations
are remarkably similar.   One of our target stars, HD 135148 has an
extreme 10830\AA\ line depth ($\sim$0.1)  likely caused by radiation from  the hot 
companion contributing to the photoionization-recombination processes
populating the lower $^3S$ level. Other stars observed in this paper
have absorption depths relative to the continuum
between 0.8 and $\sim$0.95.  Fig. 15 shows that such absorption depths 
occur for a population ratio of log  n($^3S$)/n($He_{tot}$) $< \  -$6.2. 
This provides our estimate of an upper limit to the population ratio, which in
turn translates into a  lower limit to the inferred mass loss rate 
for the discussion in Section 4.3.  

These models and the assumptions of the Sobolev approximation contain
uncertainties.  The atomic physics (collisional excitation rates, 
photoexcitation and  photoionization cross-sections, and
radiative and dielectronic recombination rates) are continually updated, and these 
calculations of populations  were made more than a year ago. Only the
giant model for $\alpha$ Boo was constructed using a metal-poor abundance set.  For the same
chromospheric input energy, one might expect the chromospheric
temperatures to be higher in a metal-poor environment because the
radiative losses are less. However the Alpha Boo (giant) model
in which the abundances are a factor of 3 less than solar appears
to give results in harmony with those from solar abundance models.
Moreover, these are 'semi-empirical' models, generally constructed
to fit observed line profiles, so that the required temperature/density
structure accommodates a non-solar metallicity.     

Our use of the Sobolev approximation introduces assumptions as well.  
We have adopted a conservative value for {\it dV/dz}. By setting this
gradient equal to $V/(R-R_\star)$ in section 4.3, our estimate
basically averages the entire velocity increase from 0 \kms\ to the 
observed \ion{He}{1} outflow velocity over the entire 1 stellar radius
of the chromosphere between the surface and the radius of \ion{He}{1} 
formation.  The wind might become more accelerated with increasing 
distance from the photosphere, [and acceleration has been observed in
red giants in globular clusters (M\'esz\'aros \etal\ 2009b)], or
else be preferentially accelerated at higher altitudes.  This would
cause our estimate of {\it dV/dz} (and hence the mass loss rate)
to be an underestimate.  We have taken the region of helium line
formation as 1 stellar radius above the photosphere, based on several
models of the  level of formation of the H$\alpha$ line placing it
at 2R$_\star$  in globular cluster red giants (Mauas \etal\ 2006; 
M\'esza\'aros \etal\ 2009a).  However, helium is formed {\it above} the
H$\alpha$ line and so this could also lead to an underestimate of the mass
loss rate.  We have set $\tau_S$ equal to 1.0 in the Sobolev
approximation, following other authors (Hartmann 1998).  It probably would be less
in the chromosphere in a fully non-LTE calculation, and thus
decrease the mass loss rate,  but other
atmospheric parameters would undoubtedly change.

As a next step, calculations of the level populations and the line
profiles should be carried out in non-LTE, with spherical coordinates, and
include velocity fields for a grid of temperatures and values of
[Fe/H]. It would be optimum also to have other chromospheric lines
than helium, such as H$\alpha$, Ca II K, and Mg II to constrain the
chromospheric structure.   While the H$\alpha$ profiles are known to
vary, currently we are ignorant of possible  changes in Ca~II, Mg~II, and the
\ion{He}{1} 10830\AA\ lines. Of course,  such
diagnostic lines must be acquired with a variety of instruments
on the ground and from space, and  realistically may be difficult
to achieve - much less achieve simultaneously.  No measure exists of X-rays from these
sources; such a measurement could also provide constraints on any 
X-ray illumination of the chromosphere.

It is reassuring that the mass loss rates for stars on the
red giant branch span values of  
$\approx 3 \times 10^{-10}$ \smpy\ to $6.3 \times 10^{-8}$\smpy,
and these values are congruent with those determined
by non-LTE spherical models of the H$\alpha$ line. Mauas \etal\ 
(2006) find values ranging from $1.1 \times 10^{-10}$ to $3.9\times
10^{-9}$ \smpy\ for 5 red giants in NGC 2808.  M\'esz\'aros \etal\
(2009a) find similar values of the mass loss rate: $5.7 \times
10^{-10}$ 
to $4.8 \times 10^{-9}$ \smpy\  for 15 red giants in 3 globular
clusters (M13, M15, and M92) from the H$\alpha$ line too.  At the highest luminosity,
the values from helium derived here appear to be higher
than the H$\alpha$ modeling, but in agreement with the 
classical 'Reimers' value (Reimers 1975).  There may well
be wind variability too, and this could cause changes
in the mass loss rate by perhaps a factor of 6, based on the
H$\alpha$ variability (M\'esz\'aros \etal\ 2009b).  Further work would be
beneficial to establish mass loss rates.

\section{Propagation of Errors in the Mass Loss Rate}

We estimate the propagated rms error  $\sigma$($\dot M$) in the mass loss rate by
evaluating the contribution of  5 variables 
to $\dot M$: {\it R}, {\it V},
$N_{rel}$= $N_1/N_H$,  $\tau_S$, and $D_V$= $dV/dz$.
Here, as defined in Section 4.3,  $R$ is the distance  from the center of the star 
with radius $R_\star$, $V$ is the expansion velocity,  $N_{rel}$ is the population of 
the lower level of the He I 10830\AA\ line relative to hydrogen,
$\tau_S$ is the Sobolev optical depth, and $D_V$ is the velocity
gradient.  The error is given by:

\begin{equation}
\sigma^2(\dot M) = \sum_{X=R, V, N_{rel},\tau_S,D_V}\left(\frac{\partial \dot M}{\partial X}\right)^2 \sigma^2(X).
\end{equation}

\noindent
By evaluating  $\left(\frac{\sigma (\dot M)}{\dot M}\right)^2$, 
$\sigma(\dot M)$ can be estimated as a function of $\dot M$ to assess the 
contribution of each quantity to the error.  Writing Equation (B1) in
terms of the variables gives,

\begin{eqnarray}
\left(\frac{\sigma(\dot M)}{\dot M}\right)^2& = & \frac{1}{\dot M^2}
 \left(\frac{\partial \dot M}{\partial N_{rel}}\right)^2
 \sigma^2(N_{rel}) + \frac{1}{\dot M^2} \left(\frac{\partial \dot
 M}{\partial \tau_S}\right)^2 \sigma^2(\tau_S) + \frac{1}{\dot M^2}
 \left(\frac{\partial \dot M}{\partial V}\right)^2 \sigma^2(V)\nonumber\\
 & + & \frac{1}{\dot M^2} \left(\frac{\partial \dot M}{\partial R}\right)^2 \sigma^2(R)+\frac{1}{\dot M^2} \left(\frac{\partial \dot M}{\partial D_V}\right)^2 \sigma^2(D_V)
\end{eqnarray}

\noindent
From Equation (5) in Section 4.3,  
\begin{equation}
\dot M = K \frac{R^2 V \tau_S  D_V}{N_{rel}}
\end{equation}

\noindent
where $K$ is a constant composed of atomic and scaling parameters. 
Taking the partial derivatives of these variables, and  reinserting the expression 
for $\dot M$, we find, 

\noindent
Substituting these quantities into Equation (B2) yields,

\begin{equation}
\left(\frac{\sigma(\dot M)}{\dot M}\right)^2 =
\left(\frac{\sigma(N_{rel})}{N_{rel}}\right)^2 +
\left(\frac{\sigma(\tau_S)}{\tau_S}\right)^2 +
\left(\frac{\sigma(V)}{V}\right)^2 
+ \left(\frac {2\sigma(R)}{R}\right)^2 +  \left(\frac{\sigma(D_V)}{D_V}\right)^2
\end{equation}

\noindent
Now supposing that $\frac{\sigma(N_{rel})}{N_{rel}} = 3$, $\frac{\sigma(\tau_S)}{\tau_S}=1$,
$\frac{\sigma(V)}{V}= \frac{1}{2}$,
$\frac{\sigma(R)}{R}=\frac{1}{2}$, and $\frac{\sigma(D_V)}{D_V}=1$ 
and substituting these values into Equation (B4), we find

\begin{equation}
\left(\frac{\sigma(\dot M)}{\dot M}\right)^2 = 9 + 1 + \frac{1}{4} + 1
+ 1
\end{equation}

\noindent
so that the largest contribution to the error arises from the uncertainty in  $N_1$, and
\begin{equation}
\sigma({\dot M}) = 3.5 \dot M.
\end{equation}

\noindent
Thus, in this Sobolev approximation, the uncertainty in the mass loss rate principally
depends on the population of the lower level of the \ion{He}{1}
10830\AA\ transition.

\clearpage  
%%%%%%%%%%%FIGURE 1
\begin{figure}
\begin{center}
\includegraphics[angle=0, scale=0.7]{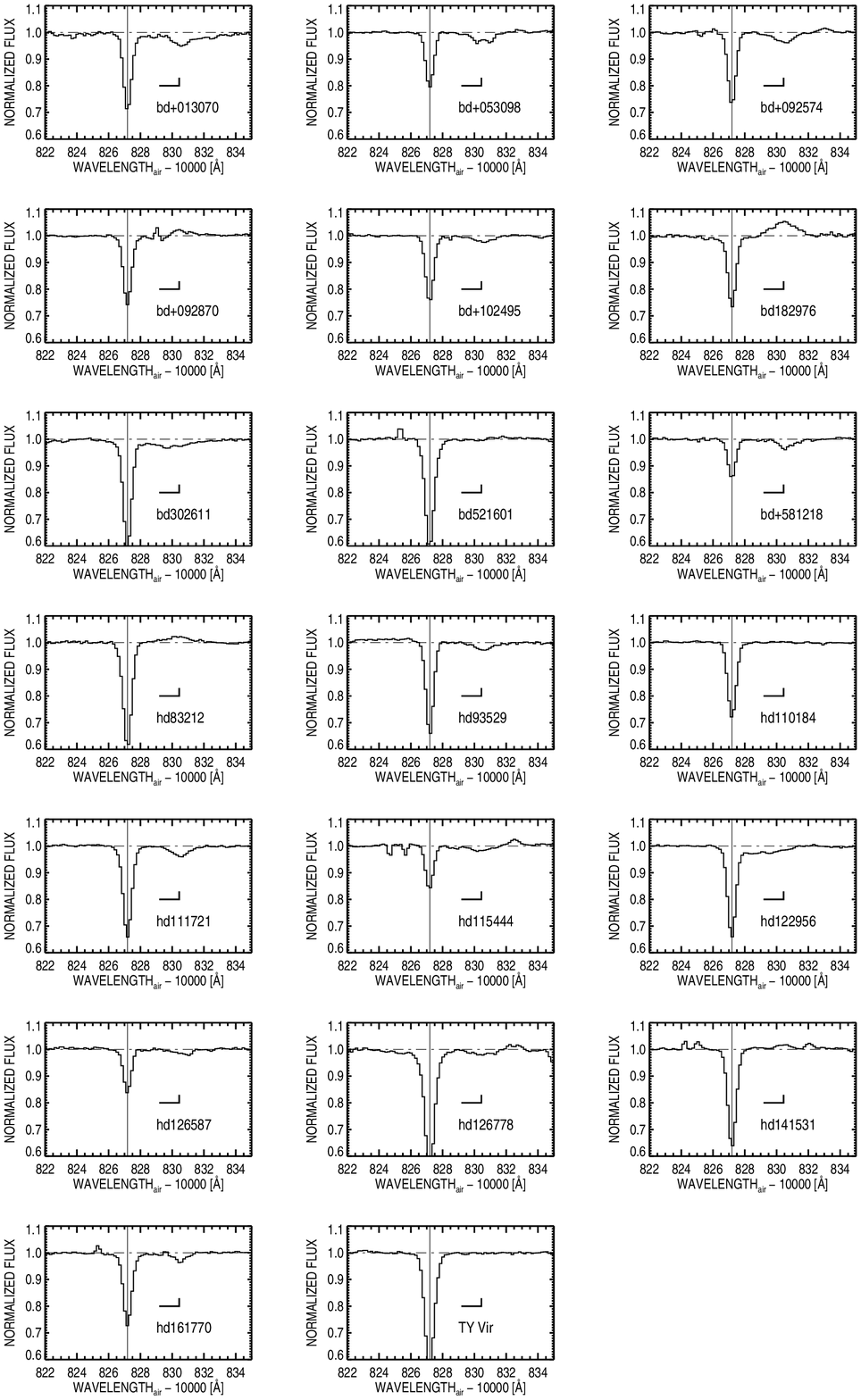}
%/data/dupree5/nirspec/new_reduction_2005/flat/all_rgb.eps
\end{center}

\vspace{-0.3in}

\caption{Red giant (RGB) stars showing the region of the He I line. 
The Si I photospheric line is the strong narrow absorption at 10827.09\AA.   
The position and extent of the He I triplet
line spanning 10830.081-10830.341\AA\ are marked.  The strongest member
of the multiplet occurs at 10830.341\AA.  Weak narrow emission features
sometimes arise from the incomplete sky subtraction. A bad pixel
causes the narrow emission spike in BD+09\arcdeg2870.}
\end{figure}

\clearpage
%%%%%%%%%%%%%FIGURE 2
\begin{figure}
\begin{center}
\includegraphics[angle=90.,scale=0.7]{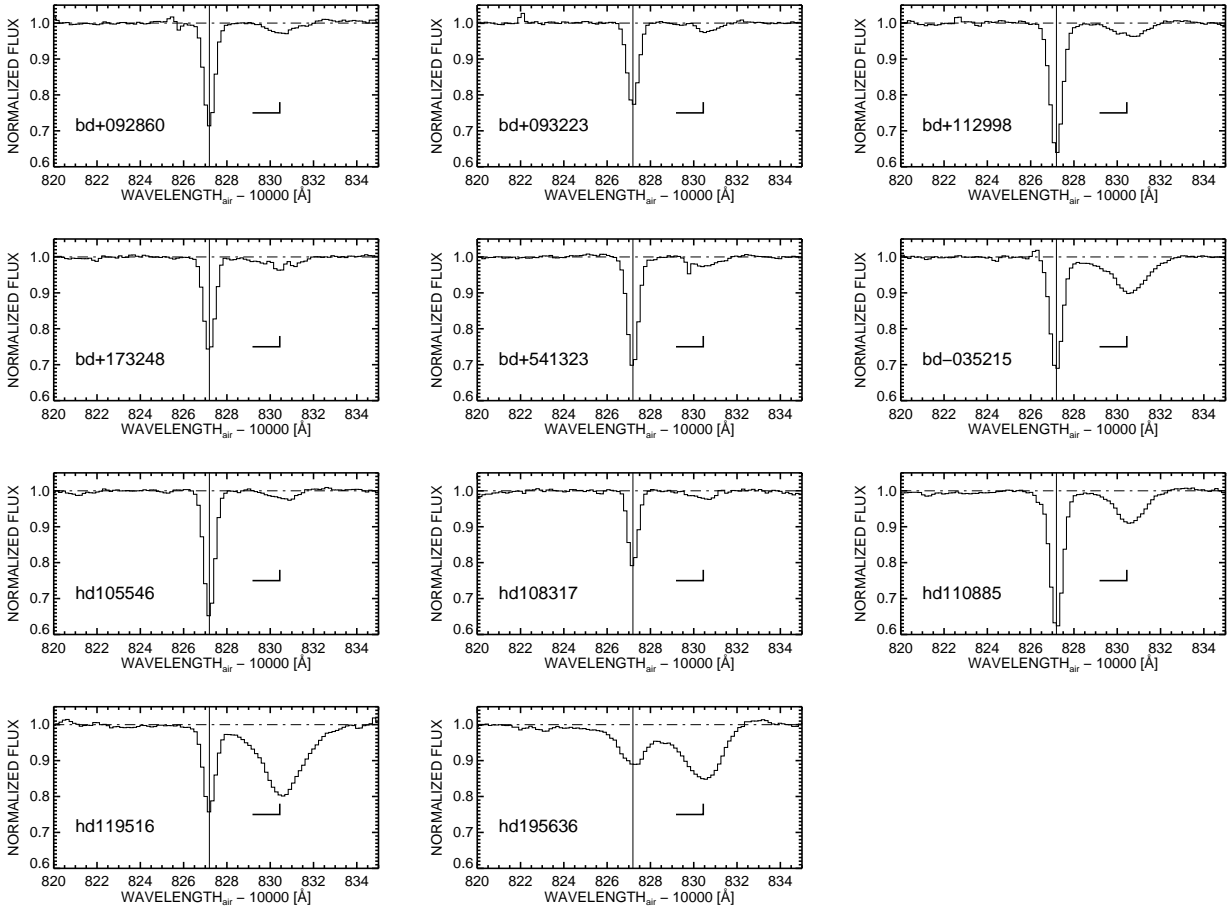}
%/data/dupree5/nirspec/new_reduction_2005/flat/all_rhb.eps}
\end{center}
\caption{Red horizontal branch stars in the sample. See caption for Figure 1}
\end{figure}

\clearpage %AGB STARS  FIG 3
%/data/dupree5/nirspec/analysis/agb_v.eps
\begin{figure}
\begin{center}
\includegraphics[scale=1.1]{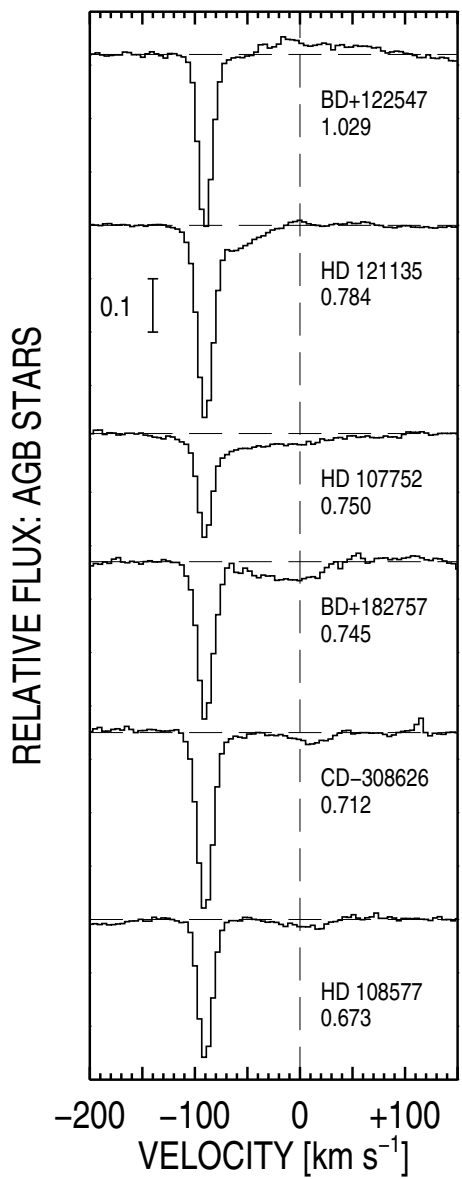}
\caption{The \ion{He}{1} line in normalized spectra of the 6 AGB stars in our
  sample. The position of the \ion{He}{1} $\lambda$10830.341 transition is marked
  by a broken line set to zero velocity. The continuum level is 
set at 1.0 for each star, with each spectrum offset by a constant
value; the extent of 0.1 in the continuum level is
shown.  The strong Si I photospheric line at $-$90 \kms\ dominates
this region of the spectrum. The AGB  stars are arranged in order 
(lower to upper spectra) 
of increasing $(B-V)_0$ values which are noted below the stellar
identification.  The most luminous
objects show weak emission in He I.  Evidence of absorption
at high negative velocities is seen in the 4 coolest stars:
BD+12\arcdeg2547, HD 121135, HD 107752, BD+18\arcdeg2757. }
\end{center}
\end{figure}
\clearpage
%%%%%FIGURE 4
%/data/dupree5/nirspec/new_reduction_2005/flat/all_sg.eps
\begin{figure}
\begin{center}
\includegraphics[scale=0.9]{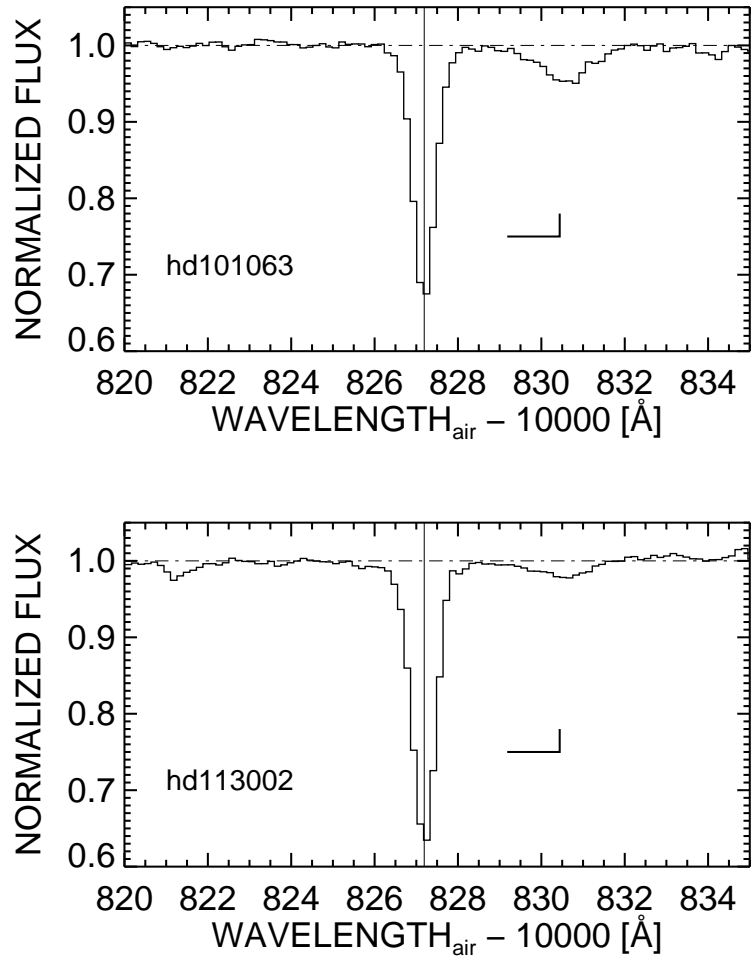}
\end{center}
\caption{He I spectra for the two subgiants in the sample.}
\end{figure}

\clearpage
%%%%%%%%%%%%%FIGURE 5
%/data/dupree5/nirspec/new_reduction_2005/flat/hd_135148.eps
\begin{figure}
\includegraphics[scale=0.7]{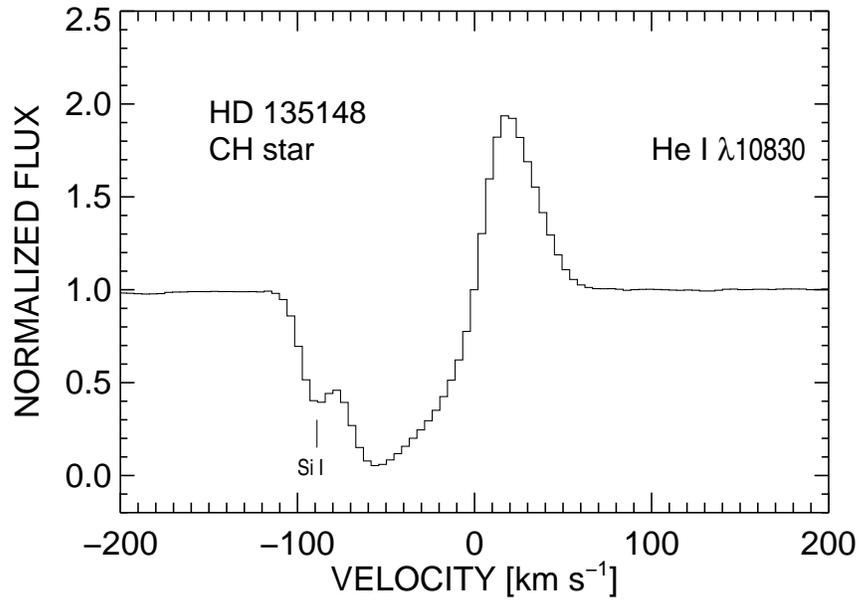}
\caption{HD 135148, a CH star, shows a substantial deep P Cygni profile with
extent at least to the Si I line and beyond that indicates a wind 
velocity of $\sim$ 115 km s$^{-1}$.}
\end{figure}

\clearpage
%%%%FIGURE 6
%/data/dupree5/nirspec/new_reduction_2005/flat/hd_104207.eps
\begin{figure}
\includegraphics[scale=0.8]{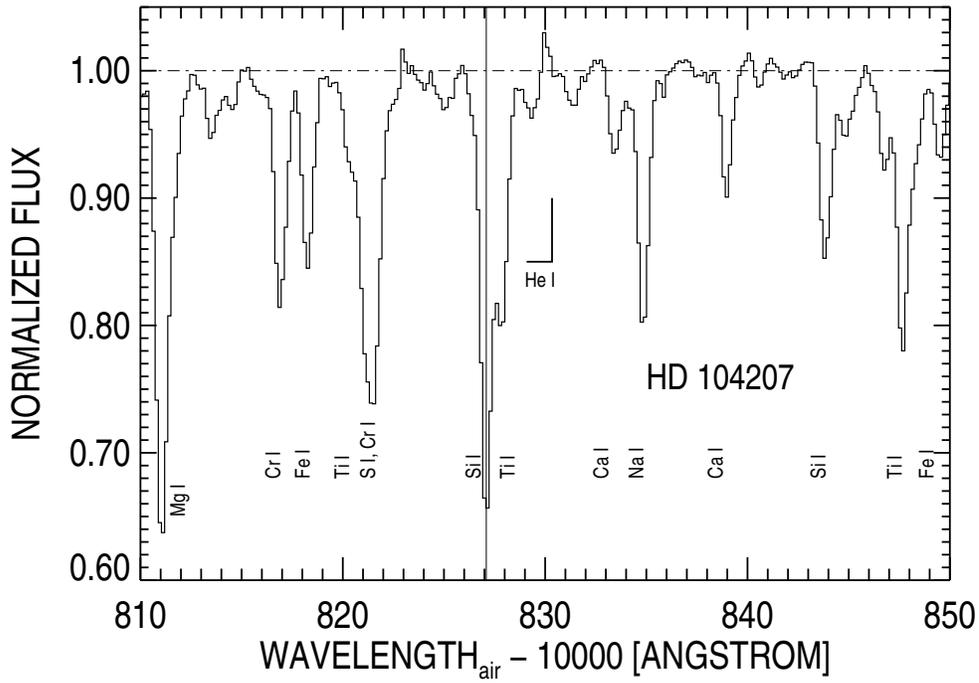}
\caption{The He spectral region of HD 104207 (GK Com), the coolest red
giant in this sample, where the spectrum is dominated by neutral lines of Si,
Ti, Fe, and Ca.  In addition, the \ion{Ti}{1} line at 10828.04\AA\  occurs 
in very cool stars near the \ion{Si}{1} transition at 10827.09\AA.  
The \ion{He}{1} line at $\lambda$10830.34  appears to have weak
emission longward of the absorption. The short
wavelength
wing of the \ion{Si}{1} 10827.09\AA\ profile may have
additional absorption when compared to the other \ion{Si}{1}
transition at 10843.9\AA, possibly caused by extended helium absorption. }
\end{figure}

\clearpage  
%%%  FIGURE 7
%/data/dupree5/nirspec/analysis/cmd.types.bw.eps
%/data/dupree5/nirspec/analysis/cmd.teff.eps
\begin{figure}
\includegraphics[angle=90,scale=0.55]{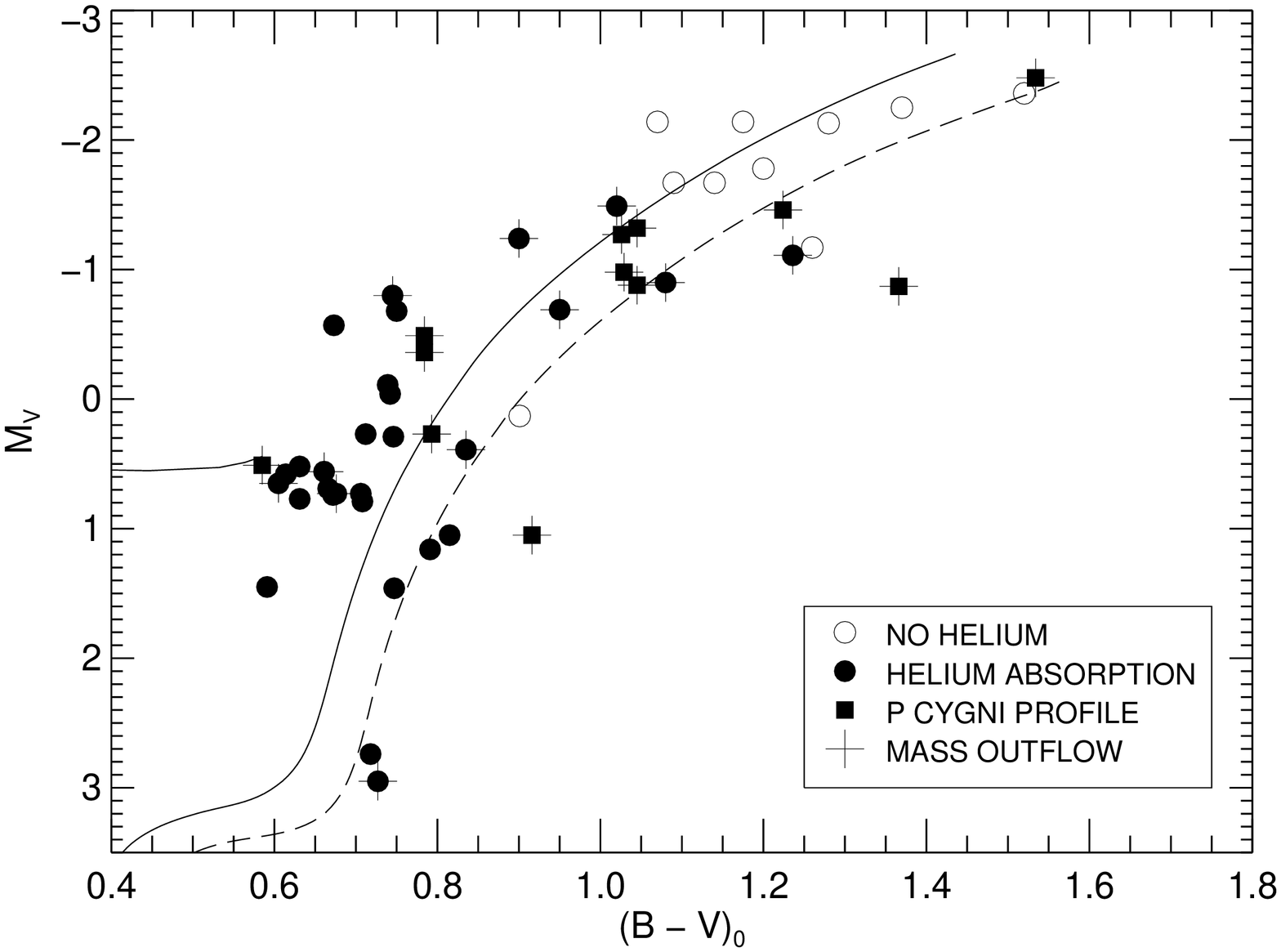}
\hspace{-0.2 in}
\includegraphics[angle=90,scale=0.55]{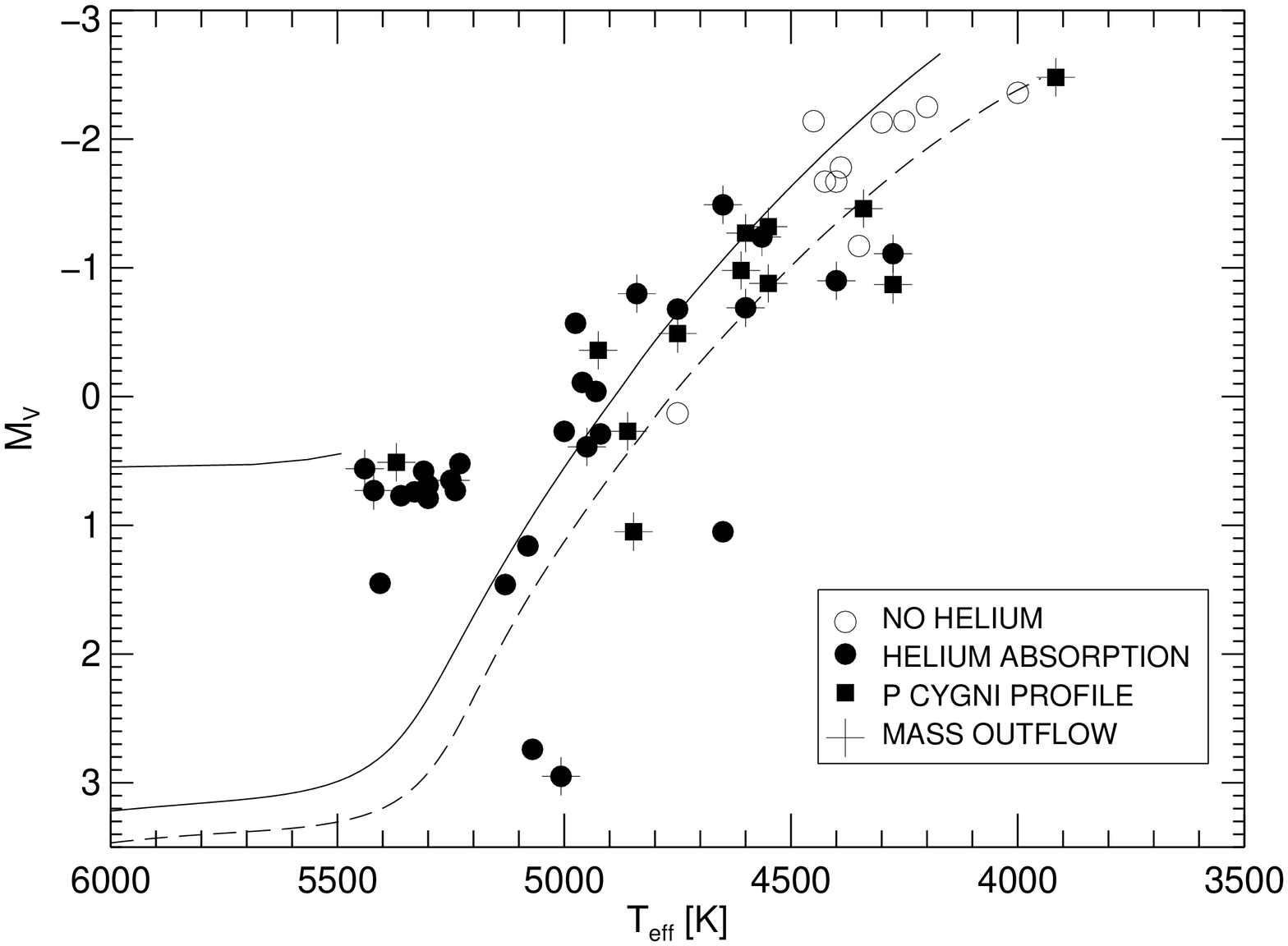}
\hspace{-0.2 in}
\caption{Location of target stars in a color-magnitude diagram ({\it
    top panel}) 
and in a T$_{eff}$-magnitude diagram ({\it lower panel}). Results
for 6 giants in M13 and 5 metal-poor field giants reported
previously (Dupree et al. 1992a; Smith et al. 2004) have been added to 
this figure (see Table 3).  The curves mark  12.1 Gyr 
isochrones (VandenBerg et al. 2006) for an abundance 
[Fe/H]=$-$2.01 ({\it solid line}) and [Fe/H]=$-$1.53 ({\it
  broken line}). Plus signs mark both P Cygni profiles and asymmetric
profiles signaling mass outflow.}
\end{figure}

\clearpage
%%FIGURE 8
%/data/dupree5/nirspec/analysis/ew_te.eps

\begin{figure}
\includegraphics[angle=90,scale=0.8]{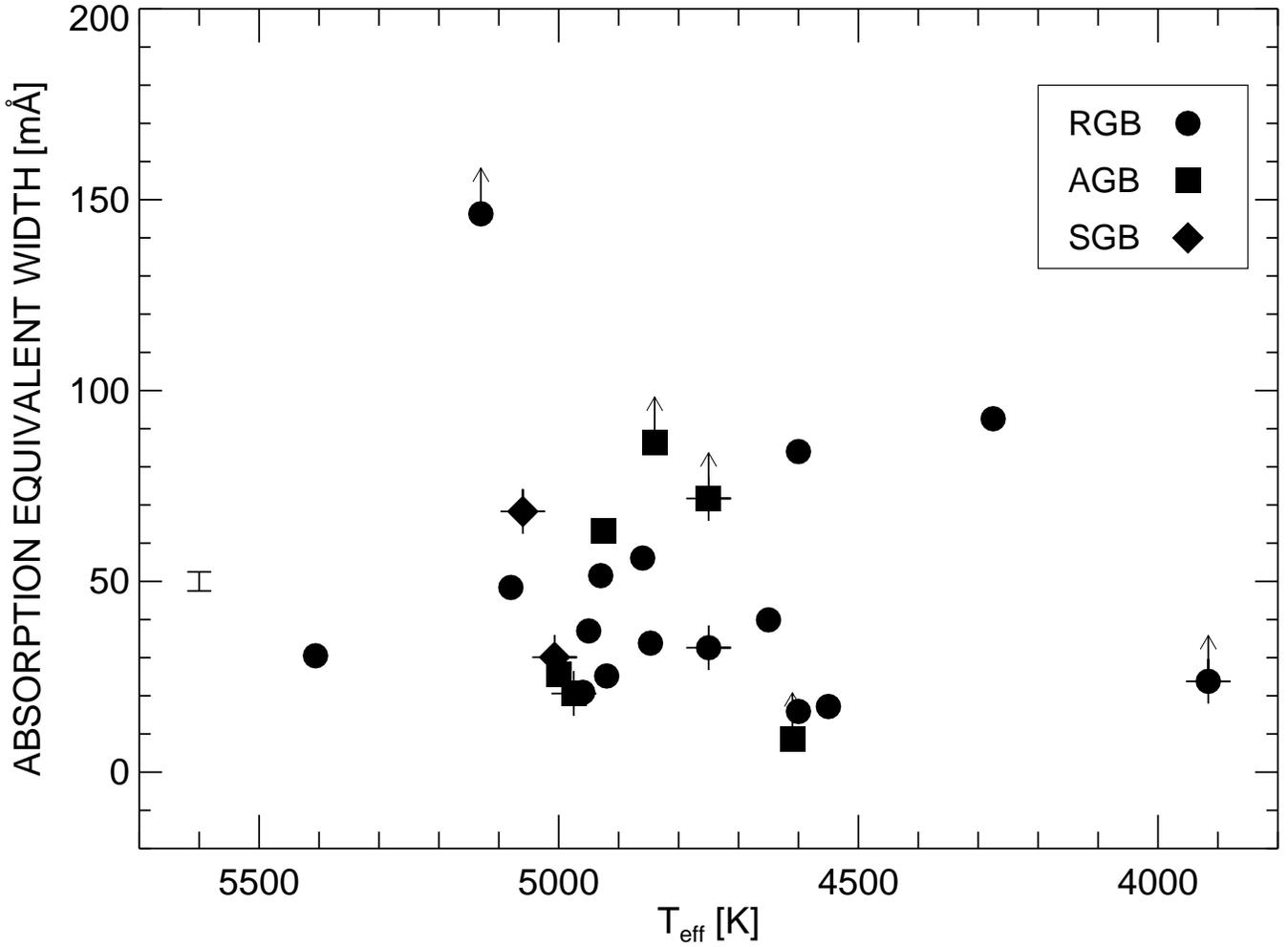}
\caption{Equivalent width of the He I absorption line as a function of
$T_{eff}$  with evolutionary status of the stars indicated as follows: 
RGB=red giant branch; AGB= asymptotic giant
branch; SGB= subgiant branch.  The plus ($+$) signs overplotted
indicate stars with P Cygni profiles. Lower limits to the equivalent
widths occur when the helium absorption overlaps the \ion{Si}{1} photospheric
line. } 

\end{figure}

\clearpage
%%FIGURE 9  rhb as f(Teff)
%/data/dupree5/nirspec/analysis/rhb_te_ew.eps
\begin{figure}
\begin{center}
\includegraphics[angle=0,scale=0.8]{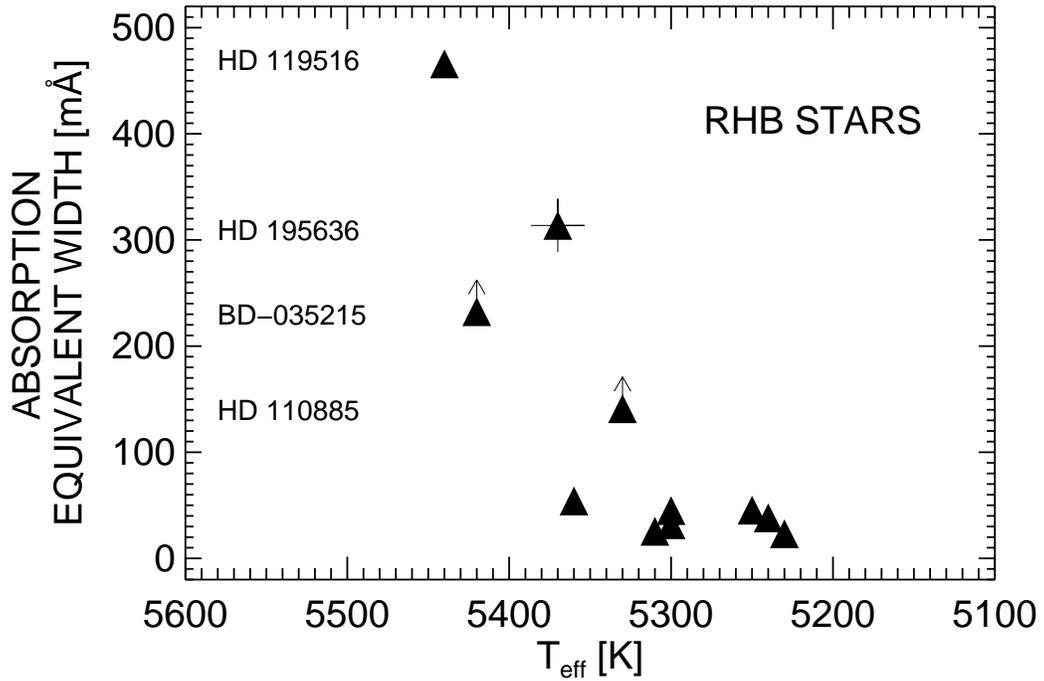}
\end{center}
\caption{Equivalent width of $\lambda$10830 in the RHB sample as a function of
  $T_{eff}$. Absorption becomes systematically stronger in stars
with $T_{eff} \ga$ 5320K.  The plus ($+$) sign overplotted
indicates a   star with a P Cygni profile. Lower limits to the equivalent
widths occur when the helium absorption overlaps the \ion{Si}{1} photospheric
line. Errors in the equivalent width amount to $\sim \pm$5\%.}
\end{figure}

\clearpage
%%%FIGURE 10 rgb for pop1 and our sample
%/data/dupree5/nirspec/analysis/ew_te_pop1.eps
\begin{figure}
\begin{center}
\includegraphics[angle=90,scale=0.8]{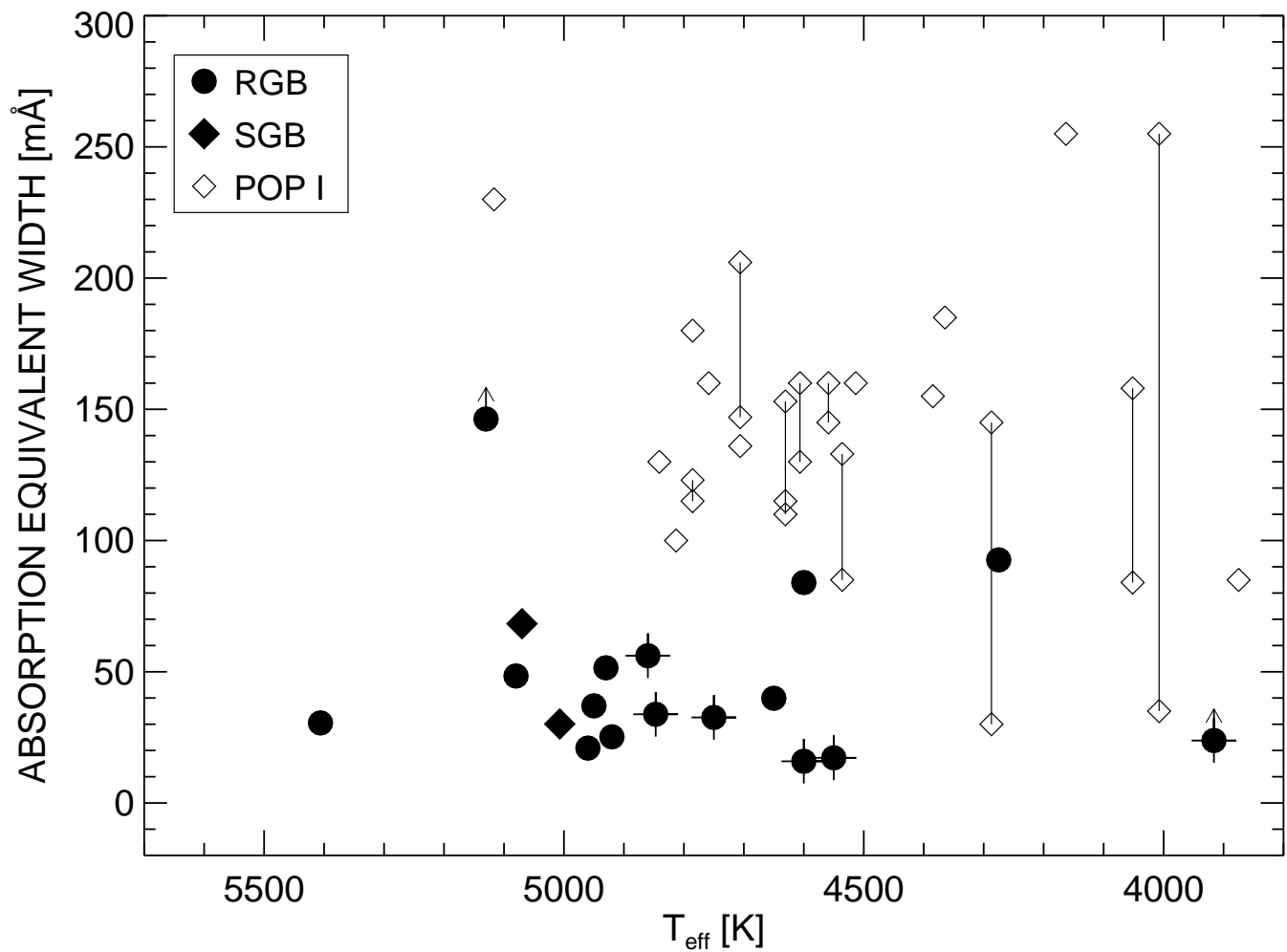}
\caption{Equivalent widths of the \ion{He}{1} absorption in red giants
  and subgiants from our sample and from the red giants in
  Population~I stars reported by  O'Brien \& Lambert (1987).  
Many of the Population I objects were observed several
  times and the lower and upper values of the equivalent widths are connected
  by solid lines.  The plus ($+$) sign overplotted indicates a P Cygni 
profile.  Some  Population I giants with  strong X-ray emission  
have substantially larger equivalent widths 
(O'Brien \& Lambert 1986; Sanz-Forcada \& Dupree
2008), and these are omitted from this figure.}
\end{center}
\end{figure}
\clearpage
%%FIGURE 11
%/data/dupree5/nirspec/analysis/fe_ew_bw.eps
\begin{figure}
\includegraphics[angle=90,scale=0.8]{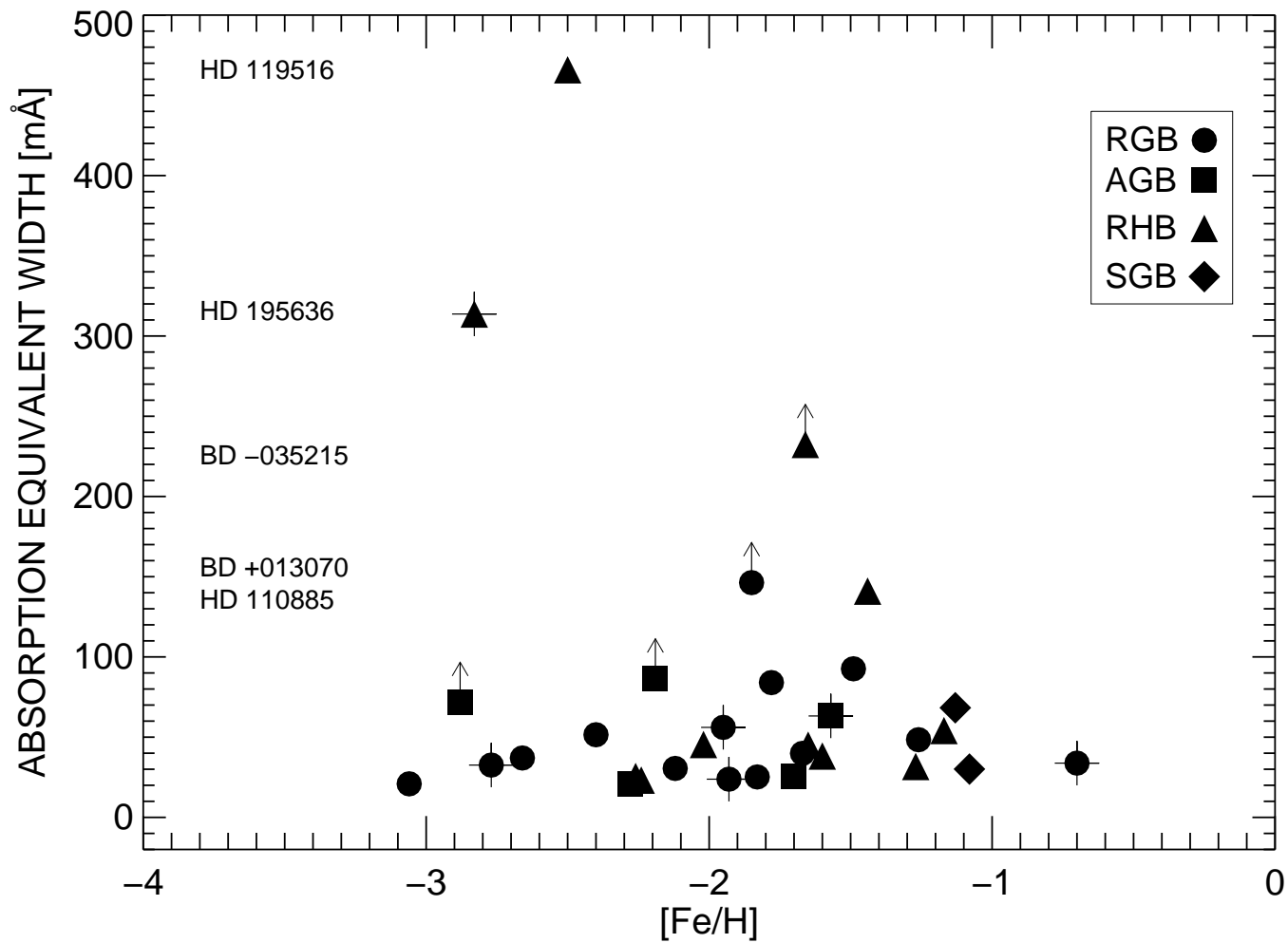}
\caption{Equivalent widths of \ion{He}{1} absorption  as a function of
  [Fe/H] with evolutionary status of the stars indicated, as follows: 
RGB=red giant branch; RHB=red horizontal branch; AGB= asymptotic giant
branch; SGB= subgiant branch.  The CH star, HD 135148 with EW= 2390 m\AA\ 
and [Fe/H]=$-$1.9 has been omitted as well as 3 stars (2 RGB and TY Vir)
  not showing helium. Stars displaying a P Cygni profile are marked with
a plus ($+$) sign; lower limits are shown where the helium absorption extends
into the neighboring \ion{Si}{1} line. 
There is no systematic dependence of the equivalent width
on [Fe/H] between [Fe/H]=$-$0.7 and $-$3.0.}
\end{figure}

%%%%% FIGURE 12
%Blue/Red velocity extent
%/data/dupree5/nirspec/new_reduction_2005/flat/ratio_mv.eps
\begin{figure}
\includegraphics[angle=90,scale=0.8]{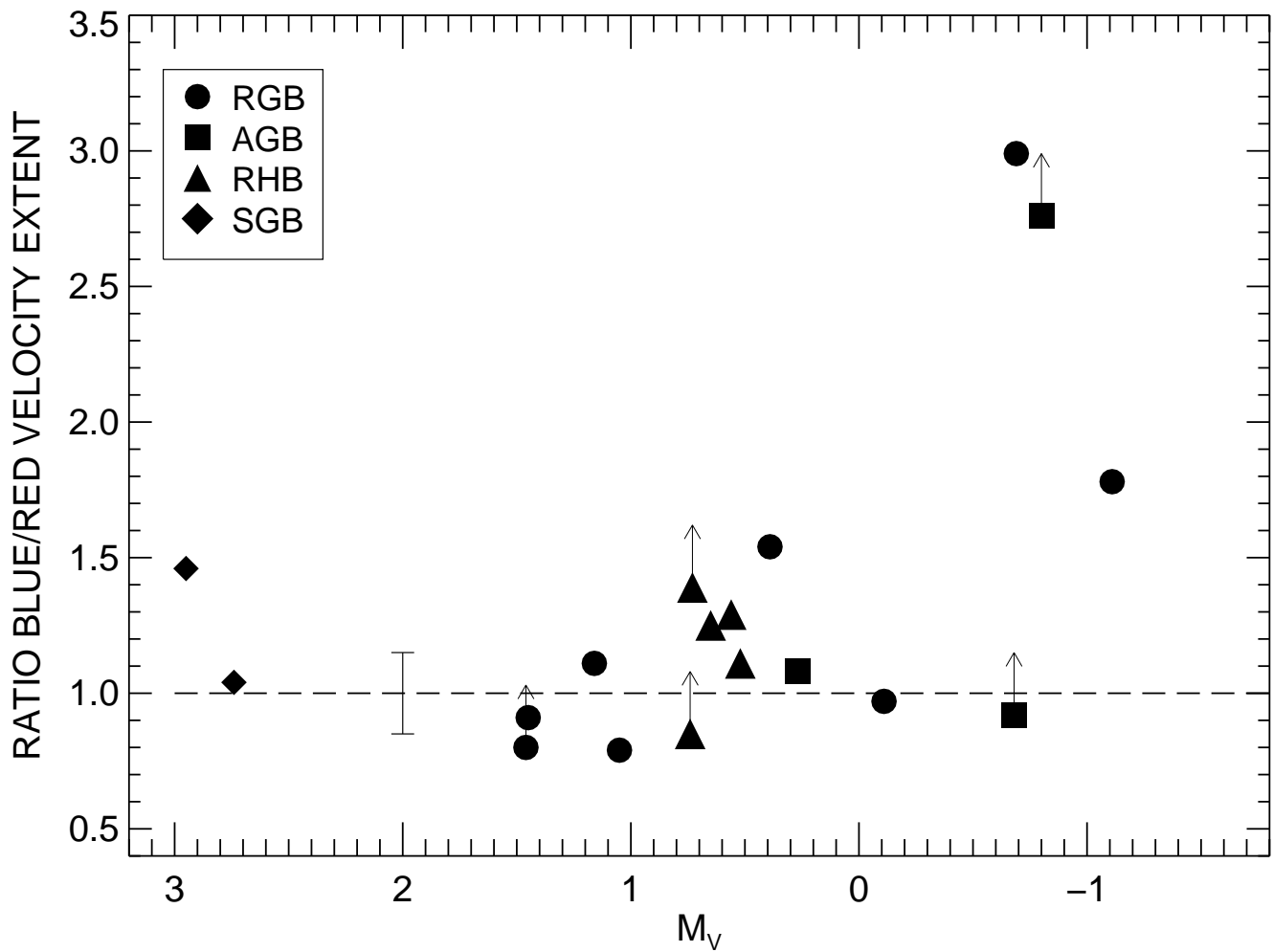}
\caption{The ratio of the short wavelength velocity extent ({\it BLUE}) to the
long wavelength velocity extent ({\it RED}) of the absorption profile
of the \ion{He}{1} 10830\AA\ line as a
function
of absolute visual magnitude. Only stars with helium absorption lines
and no emission 
are included here.   A  line arising in a stationary atmosphere has a ratio
of 1; outflow is indicated when Blue/Red $>$1.   
An 'up' arrow marks stars for which the blue velocity
wing reached the \ion{Si}{1} line, 90 \kms\ to shorter wavelengths,
but the limiting extent could not be determined because of overlap
with
the \ion{Si}{1} feature.  The error bar marks the estimated 15\% uncertainty in
the measurement of the ratio.  The  majority of
these
stars have asymmetric absorption profiles signalling outflow.}
\end{figure}

\clearpage 
%%FIGURE 13
%/data/dupree5/nirspec/analysis/feh_vel.eps
\begin{figure}
\includegraphics[angle=90,scale=0.8]{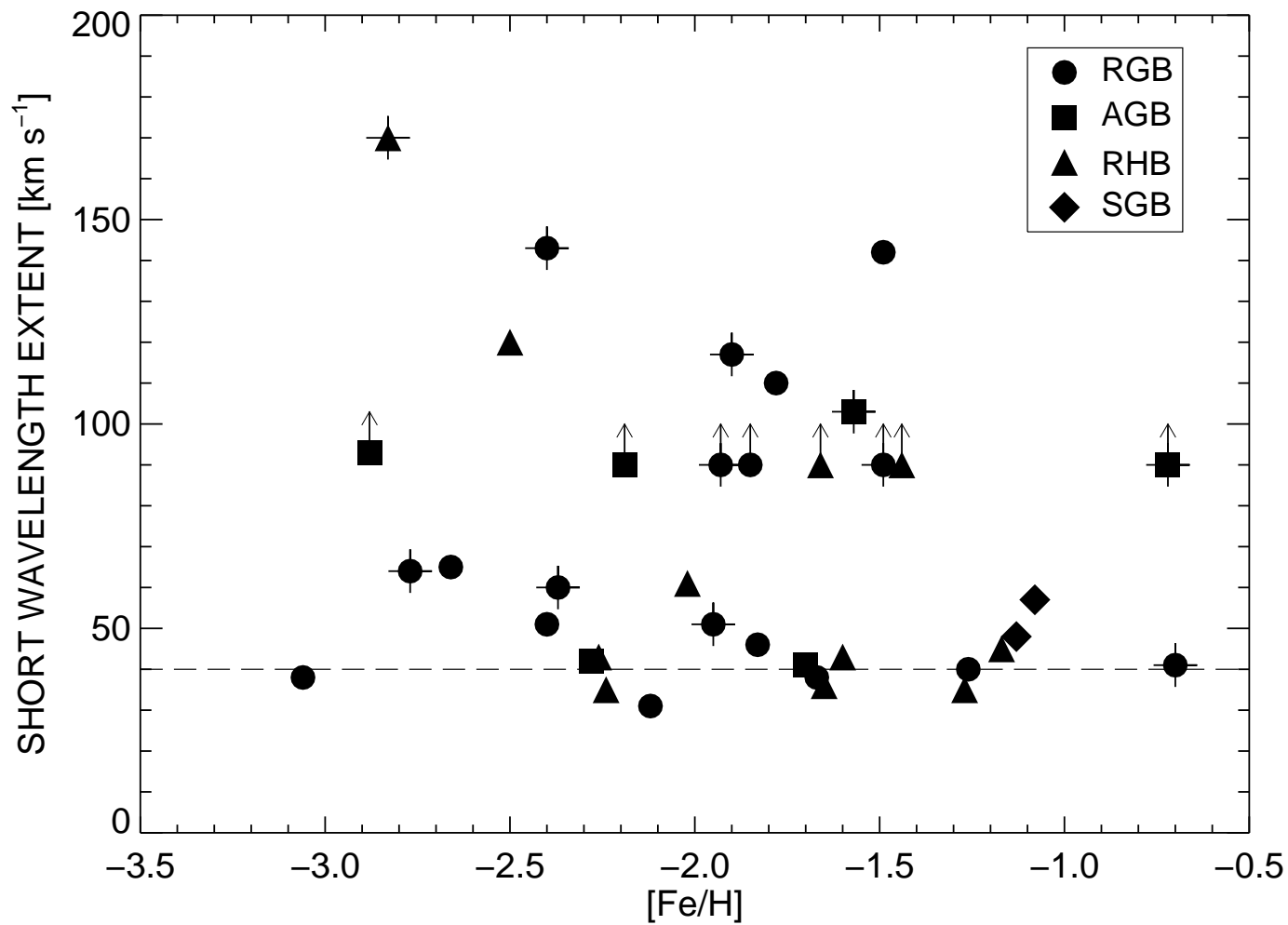}
\caption{Short wavelength extent
  of the helium 10830\AA\ line as a function of [Fe/H].  An extent of
  $\sim$40 \kms\ appears to
correspond to the normal thermal and turbulent 
width of the line (marked by the broken line). No systematic 
dependence is present  as a function of metallicity.  Stars displaying
  a P Cygni profile are marked with a plus ($+$) sign.  Upward
  pointing arrows denote lower limits because the helium absorption extends into the 
neighboring \ion{Si}{1} line and the extent of the helium profile can 
not be determined.}
\end{figure}

\clearpage
%%FIGURE 14
%/data/dupree5/nirspec/analysis/mlr_evol.eps
\begin{figure}
\includegraphics[angle=90,scale=0.8]{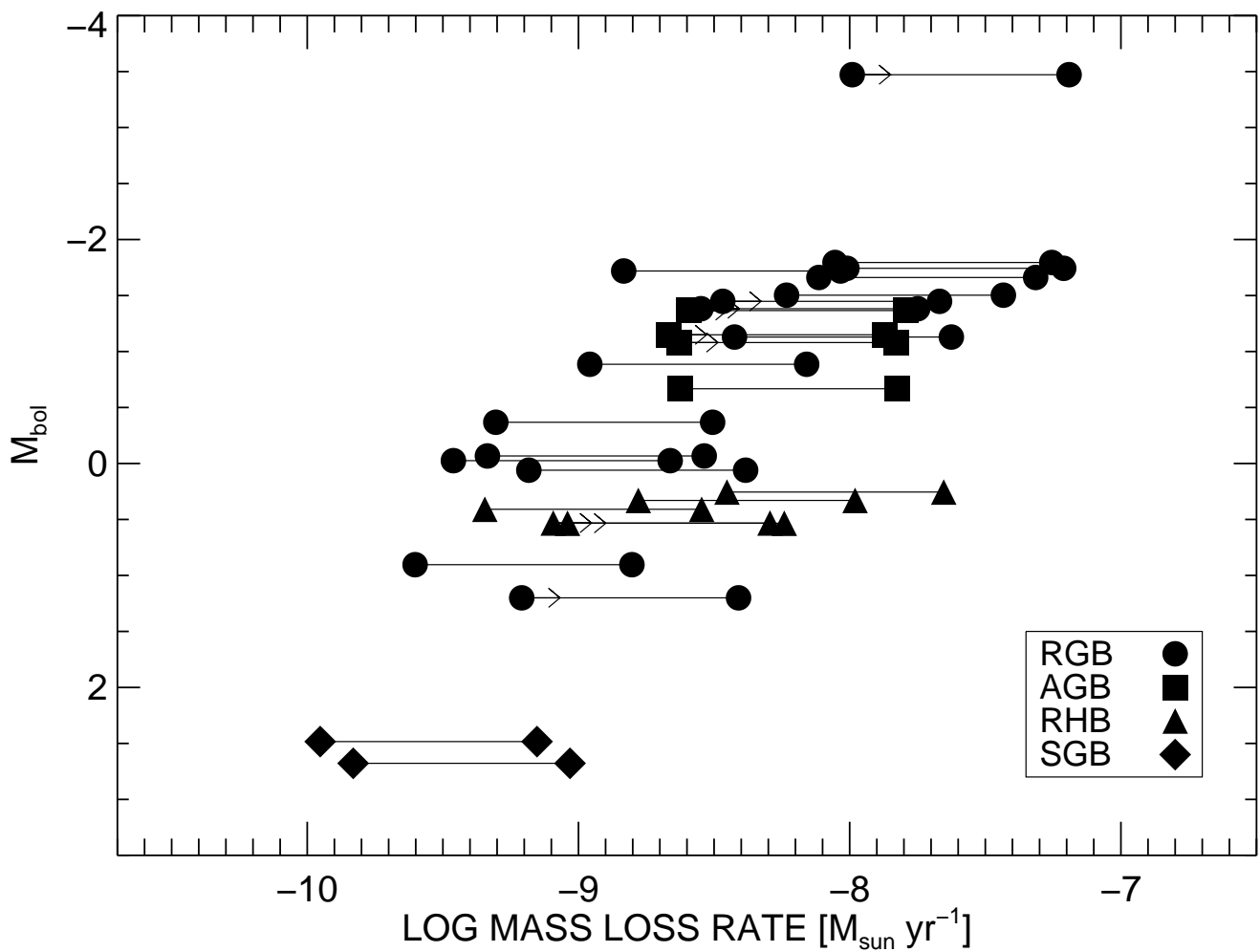}
\caption{The relation between the bolometric absolute magnitude, $M_{bol}$, and the mass
loss rate (Equation 7) as inferred from the \ion{He}{1} 10830\AA\ profile  in the
stars with $V_{term}$ exceeding the thermal width of 45 \kms.  Upper and lower limits are 
shown that derive from the
limits on the  the lower level population ratio. Some stars exhibit absorption
extending into the Si I
line located  $-$90 \kms\ from  \ion{He}{1}.  Since the
termination
of the \ion{He}{1} profile is difficult to establish, the figure
shows a lower limit arrow on these rates. }

\end{figure}

\appendix

%APPENDIX  Fig. 15
%/home/dupree/helium/keck/paper2/he_pop.eps
\begin{figure}
\begin{center}
\includegraphics[angle=0,scale=0.9]{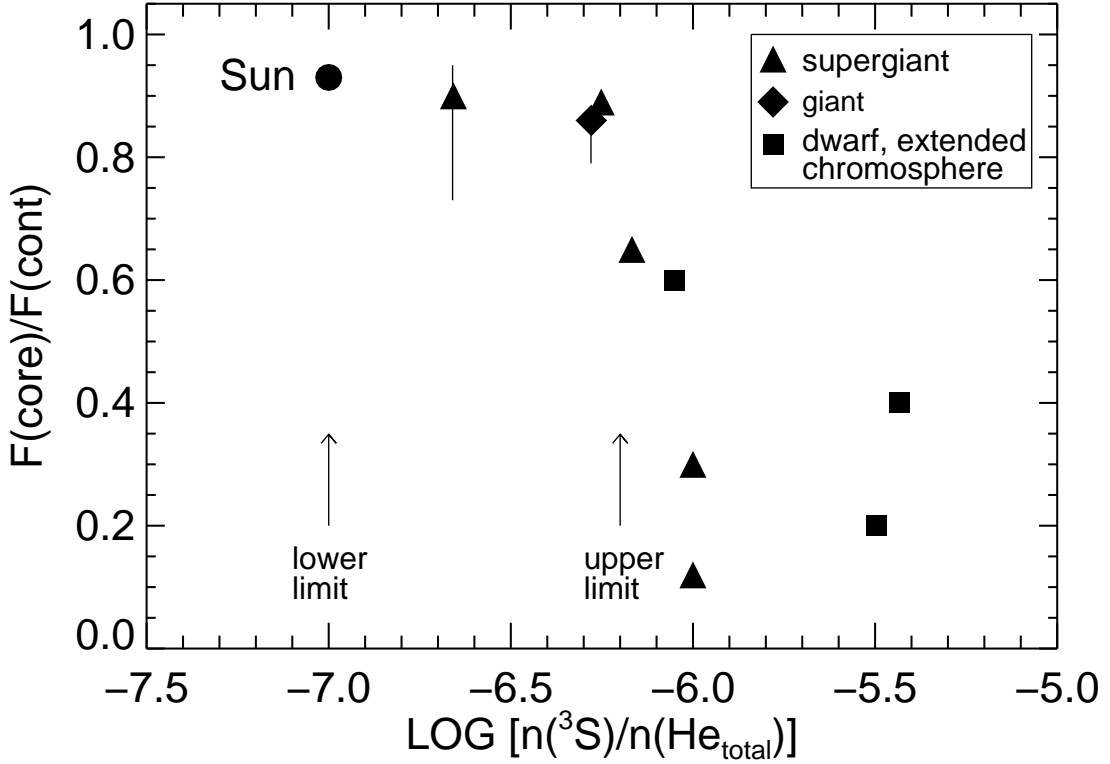}
\caption{The depth of the \ion{He}{1} 10830\AA\ absorption as
a function of the ratio of the lower level population, n($^3S$), in 
\ion{He}{1} to the
total helium abundance, n(He$_{total}$), as calculated for various stellar
chromospheres. Here, He$_{total}$= \ion{He}{1} + \ion{He}{2}.  The
vertical bars on two stars indicate the span of line-core depth values at different
$\mu$ where $\mu$=cos $\theta$, the polar angle with respect to
the stellar photosphere in the plane-parallel approximation that was
used. With the exception also of the Sun, all of the other models were 
calculated in a spherical approximation.  The giant star model
has parameters:  $T_{eff}$ = 4250K, log g = 1.6, and solar abundances
other than hydrogen and helium are decreased by a factor of 0.33.
From these
results, we take an {\it upper limit} of the lower level population
ratio, n($^3S$)/n($He_{total}$)= $-$6.2 dex, and a {\it lower limit}
of this ratio of $-$7.0 dex. } 
\end{center}
\end{figure}

%TABLE 1
%-------------------begin table
%\documentclass[12pt,preprint]{aastex}
%\begin{document}
\begin{deluxetable}{lrrrcrrrrrccrlcl}
\def\a{\phantom{0}}
\def\b{\phantom{00}}
\tablecolumns{16}
\rotate
\tabletypesize{\scriptsize}
\tablewidth{0pt}\tablenum{1}
\tablecaption{Metal-Poor Field Giants Observed}
\tablehead{
\colhead{Star}   &     
\colhead{$V$}  &     
\colhead{$B-V$}  &
\colhead{$U-B$} &
\colhead{Ref.}  &
\colhead{$E(b-y)$} &
\colhead{$M_V$} &
\colhead{$V$} &
\colhead{$b-y$} &
\colhead{$c_1$} &
\colhead{Ref.}  &
\colhead{E($B-V$)} &
\colhead{($B-V$)$_0$} &
\colhead{[Fe/H]} &
\colhead{Ref.}  &
\colhead{Evol.\tablenotemark{a}} 
 }
\startdata
BD +01 3070&  10.060& 0.769&  \nodata&    3 &   0.016&   1.46&  10.038&  0.487&  0.264 &  2&   0.022&   0.747&    $-$1.85&   2&   RGB   \\     
BD +05 3098&  10.530&  0.780&  \nodata &   3 &   0.028&  $-$0.04&  10.537&  0.542 &  0.380&    2&   0.038   &0.742&    $-$2.4&    2&   RGB   \\    
BD +09 2574 & 10.517 &  0.793 & \nodata&    3&    0.000&   0.27 & 10.523&  0.530&  0.379&   2&   0.00\a&    0.793&    $-$1.95&   2&   RGB   \\    
BD +09 2860&  10.830  &  0.710&  \nodata&    3&    0.003&   0.73&  10.83\a&   0.440&  0.430&   2&   0.004&   0.706 &   $-$1.6 &   2&   RHB   \\    
BD +09 2870&  9.440&   1.042&   \nodata &   3&    0.012&  $-$1.27 &  9.426&  0.647&  0.527&   2  & 0.016 &  1.026&    $-$2.37 &   6 &   RGB   \\    
BD +09 3223&  9.260&  0.670&  \nodata&    3 &   0.041  & 0.58&   9.253&  0.469&  0.481&   2 &  0.056&   0.614&      $-$2.26 &   6&    RHB   \\    
BD +10 2495&  9.723&   0.749&   \nodata &   3 &   0.002&    0.29&    9.745  & 0.522  & 0.361 &   2  &  0.003 &   0.746    & $-$1.83 &   7 &   RGB   \\    
BD +11 2998&  9.067 &   0.679&   \nodata&    3&    0.035&     0.77   & 9.058&   0.453&   0.505&    2&    0.048&    0.631&     $-$1.17&    6&    RHB   \\  
BD +12 2547&  9.920 &   1.034&   \nodata &   3 &   0.004&   $-$0.94 &   9.92\a &   0.635&   0.420&    2 &   0.005 &   1.029  &  $-$0.72 &  7  & AGB   \\    
BD +17 3248&  9.37\a  &   0.66\a&    0.08\a&     1 &   0.040&   0.65&   9.352&  0.486&  0.445&   2&   0.055 &  0.605 &   $-$2.02  & 6 &    RHB   \\       
BD +18 2757&  9.795 &   0.745&   0.155&    1 &   0.000&  $-$0.80&  9.84\a&  0.550& 0.490&  2&  0.00&  0.745&    $-$2.19&    6&   AGB   \\     
BD +18 2976&  9.850 &   1.051&   \nodata&    3& 0.005&   $-$1.32&   9.835&   0.655&  0.527&   2&    0.006&   1.045&   $-$2.4&   2&   RGB   \\       
BD +30 2611&  9.125 &   1.240&   1.125&    1&   0.003&   $-$1.11&    9.143&    0.807&    0.551& 2&  0.004& 1.236& $-$1.49& 6& RGB   \\    
BD +52 1601&  8.800 & 0.901&   \nodata&    3&  0.000&   0.13&     8.80\a&    0.555&   0.445&    2&    0.00 &    0.901&     $-$1.58 &   6&    RGB   \\    
BD +54 1323&  9.343 &   0.670 &  \nodata &   3&  0.003&    0.69&    9.33\a&    0.470 &  0.440&   2&   0.004&   0.666&    $-$1.65 &  6&   RHB   \\    
BD +58 1218&  9.960 &   0.835&   \nodata&    3  &   0.000&   0.39&   9.96\a &  0.515&  0.360&    2&    0.00&     0.835&     $-$2.66&    7&   RGB   \\    
BD $-$03 5215&  10.170 &   0.719&   \nodata&    3 &   0.032&   0.73&  10.188&  0.435&  0.499&    2&    0.043 &   0.676 &    $-$1.66 &  1&    RHB   \\      
CD $-$30 8626&  9.703 &   0.759&   0.250 &   1  &  0.034&   0.27&   9.719&  0.521&  0.479  &   2  &   0.047  &   0.712  &     $-$1.7 &    2&    AGB   \\     
HD 83212  &  8.335&   1.070&   0.760&  1&   0.018&  $-$0.92&  8.328&    0.694&  0.571& 2&  0.025& 1.045& $-$1.49& 2& RGB  \\
HD 93529  &  9.300&   0.881&   0.370&  1&   0.048&   1.05&  9.306&    0.582&  0.409& 2&  0.066& 0.815& $-$1.67& 6& RGB   \\ 
HD 101063 &  9.460&   0.755&   0.098&  1&   0.027&   2.74&   9.47\a&    0.499&  0.284& 2&  0.037& 0.718& $-$1.13& 7& SGB   \\ 
HD 104207 &  6.984&   1.574&   1.565&  1& \nodata&   $-$2.48&\nodata&\nodata&\nodata&  4&  0.04& 1.534&  $-$1.93& 4& RGB    \\ 
HD 105546 &  8.616&   0.708&  \nodata & 3& 0.000&    0.79&   8.61\a&  0.460&   0.420& 2&  0.00\a& 0.708&  $-$1.27& 6& RHB \\  
HD 107752 &  10.05\a&   0.75\a &      0.18\a& 1&  0.000&  $-$0.68&  9.994& 0.578&   0.463& 2&     0.00\a&  0.75  &    $-$2.88    &7  &  AGB   \\  
HD 108317 &   8.038&   0.631&  \nodata  &  3&    0.000 &   0.52&   8.044 &  0.450 &  0.311 &   2 &   0.00\a   &  0.631  &   $-$2.24  &  6   & RHB  \\   
HD 108577 &   9.597&  0.694 & 0.134   & 1 &   0.015 &   $-$0.57 &  9.581 &  0.506 &  0.500    &2  &  0.021  &  0.673  &   $-$2.28  &    6   &   AGB  \\      
HD 110184 &   8.305&   1.175&  0.765 &    1&    0.000  &  $-$2.14 &  8.293 &  0.818 &  0.712  &  2  &  0.00\a  &   1.175   &  $-$2.56  &  6  &  RGB  \\     
HD 110885 &   9.180  &  0.672 &  \nodata  &  3 &   0.000  &  0.74 &  9.18\a  &  0.423 &  0.492  &  2    &0.00\a  &   0.672 &    $-$1.44 &   8 &   RHB   \\    
HD 111721 &   7.971 &   0.799 &  0.157  &  1 &   0.006  &  1.16  &   7.98\a &  0.526 &  0.315  &  2  &  0.008 &   0.791  &   $-$1.26   & 8  &  RGB  \\     
HD 113002 &   8.745&      0.747 &  0.209  &  1 &   \nodata  &  2.95 &  \nodata  & \nodata  & \nodata  &  5    &0.02\a  &   0.727  &   $-$1.08  &  5  &  SGB   \\    
HD 115444 &   8.967&   0.784 &  0.173 &   1 &    0.000 &  $-$0.49  &  8.98\a  &  0.575 &  0.425   & 2    &0.00\a  &   0.784    & $-$2.77  &  6  &  RGB  \\     
HD 119516 &   9.090 &  0.661 &  \nodata &  3 &   0.000  &   0.56  &9.09\a  &   0.410  & 0.525 &   2 &   0.00\a  &     0.661    &$-$2.5   &    2  &   RHB  \\  
HD 121135 &   9.357&   0.795  & \nodata &  3 &   0.008  & $-$0.36 &  9.368&   0.530&  0.509&   2&   0.011 &  0.784 &     $-$1.57 &   6 &   AGB   \\    
HD 122956  &   7.22\a&      1.01\a &   \nodata &  1&    0.042  &  $-$0.69  &   7.251 &  0.667 &  0.480&   2&   0.058 &  0.95\a &    $-$1.78   &6 &  RGB   \\ 
HD 126587  &   9.125&    0.818 &  0.160 &  1 &   0.058 &  $-$0.11&   9.097&  0.596&  0.381&   2 &  0.079&   0.739 &   $-$3.06&   7&   RGB  \\     
HD 126778  &   8.168&    0.916 &  0.666 &  1 &   0.000  &  1.05&   8.15\a&   0.596&  0.449&   2&   0.00\a&    0.916&    $-$0.7&    2&   RGB  \\     
HD 135148  &   9.490 &   1.388 &  \nodata &  3 &   0.016  & $-$0.87 &  9.425&  0.869&  0.440&   2&   0.022  & 1.366 &   $-$1.90 &  6&   RGB\tablenotemark{b}    \\  
HD 141531  &   9.130&    1.240 &  \nodata &  3 &   0.012 &  $-$1.46&    9.145&   0.765&   0.603&    2&    0.016&    1.224&     $-$1.62  & 8&   RGB   \\     
HD 161770 &   9.681&    0.665 & $-$0.041  &   1 &   0.054 &   1.45 &  9.70\a&   0.500&  0.281  & 2 &  0.074 &  0.591&    $-$2.12&   7&   RGB  \\    
HD 195636 &   9.540&    0.645 & $-$0.005  &  1 &   0.044 &   0.51&   9.552&  0.467&  0.481 &     2&   0.060&   0.585&    $-$2.83&  10&   RHB  \\    
TY Vir  &     8.1\b &        1.28\a  &  1.00\a  &   1 &   0.012 &  $-$1.17&   8.165&  0.938&  0.711&   2&   0.016&   1.26\a&     $-$1.78&   9&   SR   \\      
\enddata
\tablenotetext{a}{Evolutionary state: RGB = red giant branch; SGB = subgiant branch; SR= semiregular variable; AGB = asymptotic giant
branch decided on the basis of {\it (b$-$y, c$_1$)} diagram and {$M_V$}; RHB = red horizontal branch decided on the basis of
{\it (b$-$y, c$_1$)} diagram and {$M_V$}.} 
\tablenotetext{b}{CH star.}
\tablerefs{1. Mermilliod \etal\ 1997; 2. Anthony-Twarog \& Twarog 1994; 3. {\it HIPPARCOS} Input Catalogue; 4. GK Com (Var.); M4 III; Beers et al. 2000 
data used to derive $M_V$;  5. Beers et al. (2000) data used to derive $M_V$; 6. Pilachowski et al. (1996); 7. Pilachowski et al. (1993); 
8. Gratton et al. 2000; 9. Fulbright 2000; 10. Anthony-Twarog \& Twarog 1998.}
\end{deluxetable}
%\end{document}

%------------------endtable

%TABLE 2
%--------------------begin table
%\documentclass[12pt,preprint]{aastex}
%\begin{document}
\begin{deluxetable}{lrcrrcrrrrrc}
\def\a{\phantom{0}}
\def\b{\phantom{00}}
\tablecolumns{12}
\rotate
%\tabletypesize{\scriptsize}
\tablewidth{0pt}\tablenum{2}
\tablecaption{Parameters of \ion{He}{1} 10830\AA\ Line}
\tablehead{
\colhead{Star}   &     
\colhead{Exp.\tablenotemark{a}}  &     
\colhead{Evol.\tablenotemark{b}}  &
\colhead{RV}  &
\colhead{T$_{eff}$}&
\colhead{Line\tablenotemark{c}} &
\colhead{EW\tablenotemark{d}}  &
\colhead{V$_{term}$\tablenotemark{e}} &
\colhead{$B/R$\tablenotemark{f}}&
\colhead{R$_\star$}&
\colhead{V$_{esc}$(2R$_\star$)\tablenotemark{g}}&
\colhead{Ref.} \\
\colhead{} &
\colhead{(s)} &
\colhead{} &
\colhead{(km s$^{-1}$)}&
\colhead{(K)} &
\colhead{} & 
\colhead{(m\AA)}&
\colhead{(km s$^{-1}$)}&
\colhead{}&
\colhead{(R$_\odot$)}&
\colhead{(km s$^{-1}$)}&
\colhead{(for T$_{eff}$)}
}
\startdata
BD +01 3070 & 400&       RGB &$-$329.9&5130 &      1 &   146.3& {\bf $\ga$90}&$\ga$0.80&6.4&150&1\\
BD +05 3098 &  1440&       RGB &$-$160.5&4930 &     1 &  51.5 &  51&1.19&14.3&100&1\\
BD +09 2574  &  1200 &      RGB &$-$49.8&4860 &     2&   56.1 &52&0.55&12.8 & 106&1\\
BD +09 2860  & 2160 &     RHB&$-$20.7 &5240   &   1 &  38.2 & 44&0.97&8.5&126&1\\
BD +09 2870  & 360&      RGB &$-$120.1&4600   &  2&  15.9 &   60&0.84&30.7& 69&2\\
BD +09 3223 &  720 &      RHB&67.3& 5310    &  1 &  25.5 &  44&1.08&8.9&123&1\\
BD +10 2495 &  840 &     RGB &263.2&4920  &     1 &  25.2&   46&1.11&12.3&108&1\\
BD +11 2998 &  480 &     RHB &50.7&5360  &     1 &  54.1 &  45&0.47&7.8&131&1\\
BD +12 2547 &  420&  AGB   &5.3&4610  &   2 & 8.8:& {\bf $\ga$90}&$\ga$0.86&26.0&67&1\\ 
BD +17 3248& 720 &    RHB  &$-$147.4&5250  &    1 &  45.4 &  62&1.25&8.8&123&2\\
BD +18 2757&  500 &    AGB   &$-$29.0&4840   &  1 &  86.3 & {\bf $\ga$90}&$\ga$2.76&21.3&74&1\\
BD +18 2976&  500 &    RGB  &$-$167.4 &4550 &    2  & 17.2: &{\bf 143}&2.65&32.5&67&1\\
BD +30 2611&  200 &     RGB &$-$282.8 &4275 &     1 &  92.6&{\bf 143}&1.78&35.9&63&2\\
BD +52 1601&  400  &   RGB  &$-$47.4 &4750  &   0 &  0.    &   \nodata&\nodata&14.6&99&2\\
BD +54 1323&  600 &    RHB  &$-$67.2& 5300  &   1 &  45.0 &  36&0.48&8.4&126&2\\
BD +58 1218&  600 &     RGB   &$-$305.2 &4950 &   1 &  37. &   65&1.54&11.7&111&3\\
BD $-$03 5215&  1200 &    RHB &$-$294.5 &5420  &    1 &  232.4 &{\bf $\ga$90}&$\ga$1.39&7.8&131&1\\
CD $-$30 8626&  600 & AGB &266.2&5000 & 1 &  25.6 &   41 &1.08&11.9&99&1\\
HD 83212 &   500&      RGB&108.0& 4550 & 2\tablenotemark{h} & $-$43.6&{\bf $\ga$90}&$\ga$1.42&26.9&73&2    \\
HD 93529 &   300   & RGB&145.6&4650  & 1  & 39.9&   38&0.79&10.2&119&2\\
HD 101063&   400&      SGB &182.6&5070 &       1 &  68.3  & 48&1.04&3.6&205&3 \\
HD 104207&   6  &    RGB  &35.6&3916   &    2 &  23.8 &{\bf $\ga$90}&$\ga$8.69&95.0&39&4\\
HD 105546&   240&      RHB&18.1&5300   &     1&   31.7&   35&0.99&8.0&130&2\\
HD 107752 &  600&      AGB &219.2&4750   &    1&   71.7&{\bf $\ga$90}&$\ga$0.92&21.4&73&1\\
HD 108317 &  150&      RHB&4.4&5230 &       1&   23.2&   35 &1.11&9.5&119&1\\
HD 108577&   400&      AGB&$-$112.1&4975  &     1 &  20.6&     42&1.11&17.8&80&2 \\
HD 110184&   120&      RGB&138.5&4250   &    0 &  0. & \nodata&\nodata&59.0&49&2 \\
HD 110885&   600&      RHB&$-$48.8&5330  &     1&   14.1 &{\bf $\ga$90}&$\ga$0.85&8.1&129&1 \\
HD 111721&   240&      RGB&20.5&5080    &    1 &  48.4 &   50 &1.11&7.5&138&5\\
HD 113002&   240&      SGB&$-$95.2&5007   &    1&   30.1 &   57&1.46&3.4&212&4 \\
HD 115444&   300&      RGB&$-$27.7 &4750  &    2 &  32.6 &   65&0.71&19.6&86&2 \\
HD 119516&   600&      RHB&$-$287.2 &5440  &    1&   466. &{\bf 121}&1.29&8.6&125&1\\
HD 121135&   600&      AGB&125.0&4925   &    2 &  63.2 &{\bf 104}&1.47&16.5&84&2\\
HD 122956&    70  &   RGB &165.2 &4600  &    1  &  84. & {\bf 110}&2.99&23.4&78&2\\
HD 126587 &   480&      RGB &149.3&4960 &     1 &  20.9 &   39&0.97&14.7&99&5\\
HD 126778 &  150 &      RGB &$-$138.8&4847 &     2 &  33.8 &    41&0.56&8.9&127&6 \\
HD 135148 &  320 &      RGB &$-$85.4&4275 &     2 &  2389.6 &{\bf 117}&1.51&32.2&67&2 \\ 
HD 141531 &  440 &       RGB&2.3&4340 &  2\tablenotemark{g}&$-$12.8&\nodata&\nodata&40.1&60&1\\ 
HD 161770 &  1020&       RGB  &$-$130.6&5406 &     1&   30.5&   32&0.91&5.7&159&4\\
HD 195636 &  600&         RHB &$-$257.7&5370  &    2&   313.7& {\bf 171}&1.76&9.1&122&1\\
TY Vir  &    70 &        SR&229.1&  4350   &   0 &  0.   &    \nodata&\nodata&34.8&58&8 \\
\enddata
\tablenotetext{a}{Total exposure time, usually divided into several
  nodded segments.}
\tablenotetext{b}{Evolutionary stage estimated from photometry:
RGB= red giant branch; AGB=asymptotic giant branch; RHB= red
horizontal branch; SGB= sub-giant branch; SR= semi-regular variable.}
\tablenotetext{c}{Code for the presence of \ion{He}{1} $\lambda$10830;
  0: no He line observed; 1: He absorption; 2: P Cygni profile or
  emission observed (2 stars).}
\tablenotetext{d}{Equivalent Width: Positive values indicate equivalent width of
  absorption line below the local continuum; negative values indicate
  equivalent width of the emission line.}
\tablenotetext{e}{Furthest extent of the short wavelength absorption
  edge.} 
\tablenotetext{f}{$B/R$  is the ratio of the short wavelength extent
of helium absorption ($V_{term}$) to the long wavelength extent. }
\tablenotetext{g}{Stellar mass assumed: RGB=0.75M$_\odot$; RHB =
  0.7M$_\odot$; AGB =0.6M$_\odot$; SGB=0.8M$_\odot$; and SR=0.6M$_\odot$.}
\tablenotetext{h}{Emission clearly present in 
HD 141531 but absorption not apparent.}

\tablerefs{1. Carney \etal\ 2003; 2. Pilachowski \etal\ 1996;
  3. Carney \etal\ 2008;   4. Alonso \etal\ 1999;  5. Rossi \etal\ 2005; 
6. Cenarro \etal\  2007  8. Andrievsky \etal\ 2007}

\end{deluxetable}
%\end{document}

%--------------------end table

%TABLE 
%\documentclass[12pt,preprint]{aastex}
%\begin{document}
\begin{deluxetable}{llllcrrrrc}
\tablecolumns{10}
\tablewidth{0pt}
\tablenum{3}
\tablecaption{\ion{He}{1} 10830\AA\ Observations from Previous Publications}
\tablehead{
\colhead{Star}   &     
\colhead{M$_V$}  &     
\colhead{(B$-$V)$_0$}  &
\colhead{[Fe/H]} &
\colhead{He 10830\AA} & 
\colhead{V$_{term}$\tablenotemark{a}} &
\colhead{T$_{eff}$}&
\colhead{R$_\star$}&
\colhead{V$_{esc}$(2R$_\star$)\tablenotemark{b}}&
\colhead{Ref.}\\
\colhead{} &
\colhead{} &
\colhead{} &
\colhead{} &
\colhead{} &
\colhead{(\kms)} &
\colhead{(K)} &
\colhead{(R$_\odot$)} &
\colhead{(km s$^{-1}$)}&
\colhead{}
}
\startdata
\sidehead{Field Giants}
BD +17\arcdeg3248\tablenotemark{c}&+0.65&0.605&$-$2.1&absorption&60 &4625&8.8&123&2\\
HD 6833  &$-$0.9&1.08&$-$0.91 &absorption&{\bf $\ga$90}&4400&29.6&70&1,3 \\
HD 122563&$-$1.24&0.90&$-$2.6&absorption&{\bf 140}&4625&29.6&70&2\\
HD 165195&$-$2.14&1.07&$-$2.4&not detect.&\nodata&4450&\nodata&\nodata&2\\
HD 221170&$-$1.67&1.09&$-$2.0&not detect.&\nodata&4425&\nodata&\nodata&2\\
\sidehead{Red Giants in M13}
II-33&$-$1.78&1.20&$-$1.51&not detect.&\nodata&4390&\nodata&\nodata&2\\
III-37&$-$1.67&1.14&$-$1.51&not detect.&\nodata&4400&\nodata&\nodata&2\\
III-63&$-$2.25&1.37&$-$1.51&not detect.&\nodata&4200&\nodata&\nodata&2\\
III-73&$-$2.13&1.28&$-$1.51&not detect.&\nodata&4300&\nodata&\nodata&2\\
IV-15\tablenotemark{d}&$-$1.49&1.02&$-$1.51&absorption&30&4650&32.7&59&2\\
IV-25&$-$2.36&1.52&$-$1.51&not detect.&\nodata&4000&\nodata&\nodata&2
\enddata
\tablenotetext{a}{Furthest extent of the short wavelength absorption
  edge.}
\tablenotetext{b}{Stellar mass assumed: RGB=0.75M$_\odot$; 
 RHB=0.7M$_\odot$; AGB =0.6M$_\odot$.}
\tablenotetext{c}{RHB star.}
\tablenotetext{d}{AGB star.}
\tablerefs{1. Dupree et al. 1992a; 2. Smith et al. 2004 3. Smith \&
  Dupree 1988} 

\end{deluxetable}

%\end{document}

%---------------------endtable


\newpage
\begin{thebibliography}{}
\bibitem[]{}Alonso, A., Arribas, S., \& Mart\'inez-Roger, C. 1999,
  \aaps,   140, 261


\bibitem[]{}Anderson, J. 2002, ASP Conf Ser. 265, ed. F. van Leeuwen,
  J.D. Hughes, \& G. Piotto (San Francisco: ASP), 87

\bibitem[]{}Andretta, V., \& Jones, H. P. 1997, \apj, 489, 375

\bibitem[]{}Andrievsky, S. M., Korotin, S. A., \& Martin, P. 2007,
  \aap, 464, 709

\bibitem[Anthony-Twarog \& Twarog(1994)]{ant94} Anthony-Twarog, B. J., \& 
  Twarog, B. A. 1994, \aj, 107, 1577
\bibitem[Anthony-Twarog \& Twarog(1998)]{ant98} Anthony-Twarog, B. J., \& 
  Twarog, B. A. 1998, \aj, 116, 1922

\bibitem[]{}Avrett, E. H. 1992, in Proceedings of the Workshop on the
  Solar Electromagnetic Radiation Study for Solar Cycle 22,
  ed. R. F. Donnelly (Virginia: National Technical Information
  Service), 20

\bibitem[Avrett \& Loeser(2003)]{avrett03} Avrett, E. H., \& Loeser,
  R. 2003, in IAU Symp 210, Modeling of Stellar Atmospheres, ed. N. Piskunov,
W. W. Weiss, \& D. F. Gray (San Francisco: ASP), A-21

\bibitem[]{}Avrett, E. H., \& Loeser, R. 2008, \apjs, 175, 229
\bibitem[]{}Baliunas, S. L., Donahue, R. A., Soon, W., \& Henry,
  G. W. 1998, in The Tenth Cambridge Workshop on Cool Stars, Stellar
Systems, and the Sun, ASP Conf. Ser. 154, ed. R. A. Donahue \&
  J. A. Bookbinder (San Francisco: ASP), 153
\bibitem[]{}Baliunas, S. L., Hartmann, L., \& Dupree, A. K. 1983, \apj,
  271, 672
\bibitem[]{}Barmby, P., Boyer, M. L., Woodward, C. E., Gehrz, R. D., 
van Loon J. Th., Fazio, G. G., Marengo, M., \& Polomski, E. 2009, AJ,
137, 207
\bibitem[]{}Bedin, L. R., Piotto, G., Anderson, J., Cassisi, S., King,
  I. R., Momany, Y., \& Carraro, G. 2004, \apj, 605, L125
\bibitem[]{}Bedin, L. R., Salaris, M., Piotto, G., Cassisi, S.,
  Milone, A. P., Anderson, J., \& King, I. R. 2008, \apj, 679, L29
\bibitem[]{}Beers, T. C., Chiba, M., Yoshii, Y., Platais, I., Hanson,
  R. B., Fuchs, B., \& Rossi, S. 2000, \aj, 119, 2866
\bibitem[]{}Bianchi, L., Ford, H., Bohlin, R., Paresce, F. \& de
  Marchi, G. 1995, \aap, 301, 537
\bibitem[Bond (1980)]{}Bond, H. E. 1980, \apjs, 44, 517
\bibitem[]{}Bowen, G. H., \& Willson, L. A. 1991, \apj, 375, L53
\bibitem[]{}Boyer, M. L., McDonald, I., van Loon, J. Th., Woodward,
  C. E., Gehrz, R. D., Evans, A., \& Dupree, A. K. 2008, AJ, 135, 1395

\bibitem[Boyer et al.(2006)]{boy06}Boyer, M. L., Woodward, C. E., van
  Loon, J. Th., Gordon, K. D., Evans, A., Gehrz, R. D., Helton,
  L. A., \& Polomski, E. F. 2006, \aj, 132, 1415

\bibitem[]{}Buonanno, R., Corsi, C. E., Pecci, F. F., Richer, H. B.,
\& Fahlman, G. G. 1993, \aj, 105, 184
\bibitem[]{}Buonanno, R., Corsi, C. E., Pulone, L., Fusi Pecci, F., \&
  Bellazzini, M. 1998, \aap, 333, 505
\bibitem[]{}Busso, M., Wasserburg, G. J., Nollett, K. M., \& Calandra,
  A. 2007, \apj, 671, 802
\bibitem[]{}Cacciari, C., et al. 2004, \aap, 413, 343
\bibitem[]{}Caloi,  V., \& D'Antona, F. 2008, \apj, 673, 847
\bibitem[]{}Carney, B. W., Gray, D. F., Yong, D., Latham, D. W.,
  Manset, N., Zelman, R., \& Laird, J. B. 2008, \aj, 135, 892

\bibitem[]{}Carney, B. W., Latham, D. W., Stefanik, R. P., Laird,
J. B., \& Morse, J. A. 2003, AJ, 125, 293.

\bibitem[]{}Castellani, V., Giannone, P., \& Renzini, A. 1971, \apss,
  10, 340

\bibitem[]{}Catelan, M., Bellazzini, M., Landsman, W. B., Ferraro,
  F. R., Fusi Pecci, F., \& Galleti, S. 2001, \aj, 122, 3171
\bibitem[]{}Cenarro, A. J. \etal\ 2007, \mnras, 374, 664
\bibitem[]{}Cox, D. 2005,  \araa, 43, 337

\bibitem[]{}Cranmer, S. 2008, \apj, 689, 316

\bibitem[]{}D'Antona, F., Caloi, V., Montalb\'an, J., Ventura, P., \& 
Gratton, R. 2002, A\&A, 395, 69

\bibitem[]{}Demarque, P., \& Eder, J.-A. 1985, in Horizontal-Branch
and UV-Bright Stars, ed. A. G. Davis Phillip (Schenectady: Davis), 91

\bibitem[]{}Dixon, W. V. D., \& Sankrit, R. 2008, \apj, 686, 1162 

\bibitem[]{}Dotter, A. 2008, \apj, 687, L21

\bibitem[]{}Dupree, A. K., Avrett, E. H., Brickhouse, N. S., Cranmer, S. R., \&
Szalai, T. 2008, Cool Stars, Stellar Systems, and the Sun: CS 14
ASP Conf. Ser. 384, ed. G. T. van Belle (San Francisco: ASP), sub77
(CD-ROM) (also astro-ph/0702395)

\bibitem[]{}Dupree, A. K., Brickhouse, N. S., Smith, G. H., \&
  Strader, J. 2005, \apj, 625, L131

\bibitem[]{}Dupree, A. K., Hartmann, L., \& Avrett, E. H. 1984,
  \apj, 281, L37

\bibitem[]{}Dupree, A. K., Li, T. Q., \& Smith, G. H. 2007, AJ, 134, 1348

\bibitem[]{}Dupree, A. K., Sasselov, D. D., \& Lester, J. B. 1992a,
  \apj, 387, L85

\bibitem[]{}Dupree, A. K., \& Smith, G. H. 1995, AJ, 110, 405

\bibitem[]{}Dupree, A.K., \& Whitney, B. A., \& Avrett, E. H. 1992b in 
Cool Stars, Stellar Systems, and the Sun: Proceedings of the Seventh Cambridge
Workshop, ASP Conf. Ser. 26, ed. M. S. Giampapa \& J. A. Bookbinder (San Francisco:
ASP), 525


\bibitem[]{}Evans, A., Stickel, M., van Loon, J. Th., Eyres, S. P. S.,
Hopwood, M. E. L., \& Penny, A. J. 2003, \aap, 408, L9

\bibitem[]{}Faulkner, D. J., \& Freeman, K. C. 1977, \apj, 211, 77

\bibitem[]{}Faulkner, D. J., Scott, T. R., Wood, P. R., \& Wright, 
A. E. 1991, \apj, 374, L45

\bibitem[]{}Frank, J., \& Gisler, G. 1976, \mnras, 176, 533

\bibitem[]{}Freire, P. C., Kramer, M., Lyne, A. G., Camilo, F.,
Manchester, R. N., \& D'Amico, N. 2001, \apj, 557, L105



\bibitem[]{}Fulbright, J. P. 2000, \aj, 120, 1841

\bibitem[]{}Girardi, L., Bressan, A., Bertelli, G., \& Chiosi, C.
2000, \aaps, 141, 371

\bibitem[]{}Goldberg, L. 1939, ApJ, 89, 673

\bibitem[]{}Gratton, R. G.,  Sneden, C., Carretta, E., \& Bragaglia, A.
2000, \aap, 354, 169

\bibitem[]{}Groenewegen, M. A. T., \& deJong, T. 1993, A\&A, 267, 410

\bibitem[]{}Hansen, B. M. S. \etal\  2007, \apj, 671, 380

\bibitem[]{}Harris, W. E. 1996, \aj, 112, 1487
\bibitem[]{}Hartmann, L. 1998, in Accretion Processes in Star Formation, 
(New York: Cambridge University Press), Ch. 8
\bibitem[]{}Harvey, J. W., \& Sheeley, N. R., Jr. 1979, Sp.Sci.Rev., 23, 139
\bibitem[]{}Ita, Y., et al. 2007, \pasj, 59, S437
\bibitem[]{}Kalirai, J. S., Bergeron, P., Hansen, B. M. S., Kelson,
 D. D., Reitzel, D. B., Rich, R. M., Richer, H. B. 2007, \apj, 671, 748

\bibitem[]{}Kayser, A., Hilker, M., Grebel, E. K., \& Willemsen,
  P. G., 2008, \aap, 486, 437

\bibitem[]{}Koopmann, R. A., Lee, Y.-W., Demarque, P., \& Howard, J. M.
1994, \apj, 423, 380

\bibitem[Lambert(1987)]{lam87} Lambert, D. L. 1987, \apjs, 65, 255

\bibitem[]{}Lamers, H. J. G. L. M., \& Cassinelli, J. P. 1999 in
{\it Introduction to Stellar Winds}, (New York: Cambridge University Press), Ch. 8

\bibitem[]{}Lee, Y.-W., Demarque, P., \& Zinn, R. 1994, \apj, 423, 248

\bibitem[]{}Lee, Y.-W., Joo, J.-M., Sohn, Y.-J., Rey, S.-C., Lee,
  H.-C., \& Walker, A. R. 1999, Nature, 402, 55


\bibitem[]{}Marshall, J. R., vanLoon, J. Th., Matsuura, M., Wood,
  P. R., Zijlstra, A. A., \& Whitelock, P. A. 2004, MNRAS, 355, 1348.
\bibitem[]{}Matsunaga, N. et al. 2008, \pasj, 60, S415

\bibitem[]{}Matt, S., \& Pudritz, R. E. 2008, \apj, 678, 1109

\bibitem[]{}Mauas, P. J. D., Cacciari, C., \& Pasquini, L. 2006,
  A\&A, 454, 609

\bibitem[]{}McDonald, I., \& van Loon, J. Th. 2007, \aap, 476, 1261

\bibitem[]{}McLaughlin, D. E., \& van der Marel, R. P. 2005,  \apjs,
  161, 304

\bibitem[]{}McLean, I. S., et al. 1998, SPIE, 3354, 566

\bibitem[]{}McLean, I. S., Graham, J. R., Becklin, E. E., Figer,
  D. F., Larkin, J. E., Levenson, N. A., \& Teplitz, H. I. 2000, SPIE,
  4008, 1048

\bibitem[]{}McLean, I. S., McGovern, M. R.,  Burgasser,
  A. J.,  Kirkpatrick, J. D., Prato, L., \& Kim, S. S.  2003, \apj, 596, 561

\bibitem[]{}Mermilliod, J.-C., Mermilliod, M., \& Hauck, B. 1997,
  \aaps, 124, 349

\bibitem[]{}M\'esz\'aros, Sz., Avrett, E. H., \& Dupree, A. K. 2009a,
  \aj, 138, 615 

\bibitem[]{}M\'esz\'aros, Sz., Dupree, A. K., \& Szalai, T. 2009b, \aj,
  137, 4282

\bibitem[]{} M\'esz\'aros, Sz., Dupree, A. K., \& Szentgyorgyi, A. 2008, 
\aj, 135, 1117


\bibitem[]{}Moehler, S., Koester, D., Zoccali, M., Ferraro, F. R.,
  Heber, U., Napiwotzki, R., \& Renzini, A. 2004, \aap, 420, 515

\bibitem[]{}Nordhaus, J., Busso, M., Wasserburg, G. J., Blackman,
  E. G., \& Palmerini, S. 2008, \apj, 684, L29

\bibitem[O'Brien \& Lambert(1986)]{obr86} O'Brien, G. T., Jr., \& Lambert, D. L.
  1986, \apjs, 62, 899 


\bibitem[]{}Okada, Y., Kokubun, M., Yuasa, T., \& Makishima, K. 2007,
  PASJ, 59, 727


\bibitem[Origlia et al.(2002)]{ori02} Origlia, L., Ferraro, F. R.,
  Fusi Pecci, F., \& Rood, R. T. 2002, \apj, 571, 458
\bibitem[Origlia et al.(2007)]{ori07} Origlia, L., Rood, R. T., Fabbri, S., 
Ferraro, F. R., Fusi Pecci, F., \& Rich, R. M. 2007, \apj, 667, L85

\bibitem[]{}Pancino, E., Ferraro, F. R., Bellazzini, M., Piotto, G.,
  \& Zoccali, M. 2000, ApJ, 534, L83

\bibitem[]{}Peterson, R. C., Rood, R. T., \& Crocker, D. A. 1995,
  \apj, 453, 214

\bibitem[]{}Pilachowski, C. A., Sneden, C., \& Booth, J. 1993, \apj,
  407, 699
\bibitem[Pilachowski et al.(1996)]{pil96} Pilachowski, C. A., Sneden, C., \&
  Kraft, R. P. 1996, \aj, 111, 1689
\bibitem[]{}Piotto, G., et al. 2007, \apj, 661, L53
\bibitem[]{}Preston, G. W., 1997, \aj, 113, 1860

\bibitem[]{}Recio-Blanco, A., Aparicio, A., Piotto, G., de Angeli, F.,
  \& Djorgovski, S. G. 2006, \aap, 452, 875
\bibitem[]{}Reimers, D. 1975, Mem. Soc. R. Sci. Liege, Ser. 6, 8, 369

\bibitem[]{}Renzini, A. 1981, in {\it Phys. Proc. in Red Giants},
  ed. I. Iben 
\& A. Renzini, (Dordrecht: Reidel), 431

\bibitem[]{}Richardson, J. D., Kasper, J. C., Wang, C., Belcher,
  J. W., \& Lazarus, A. J. 2008, Nature, 454, 63
\bibitem[]{}Richer, H. B., \etal\ 2008, \aj, 135, 2141
\bibitem[]{}Rood, R. T. 1973, \apj, 184, 815
\bibitem[]{}Rossi, S., Beers, T. C., Sneden, C., Sevastyanenko, T.,
  Rhee, J., \& Marsteller, B. 2005, \aj, 130, 2804
\bibitem[]{}Sandage, A., \& Wildey, R. 1967, \apj, 150, 469
\bibitem[]{}Sandquist, E. L., \& Martel, A. R. 2007, ApJ, 654, L65
\bibitem[]{}Sanz-Forcada, J., \& Dupree, A. K. 2008, A\&A, 488, 715
\bibitem[]{}Sasselov, D. D., \& Lester, J. B. 1994, \apj, 423, 795
\bibitem[]{}Schwarzschild, M. 1970, Quart, J. R. A. S. 11, 12

\bibitem[]{}Sills, A., \& Pinsonneault, M. H. 2000, \apj, 540, 489
\bibitem[]{}Smith, G. H. 1998, PASP, 110, 1119
\bibitem[]{}Smith, G. H. 1999, PASP, 111, 980
\bibitem[]{}Smith, G. H., \& Dupree, A. K. 1988, \aj, 95, 1547
\bibitem[]{}Smith, G. H., Dupree, A. K., \& Strader, J. 2004, PASP,
  116, 819

\bibitem[]{}Sneden, C., Kraft, R. P., Guhathakurta, P., Peterson,
  R. C., \& Fulbright, J. P. 2004, \aj, 127, 2162

\bibitem[]{}Soker, N., Catelan, M., Rood, R. T., \& Harpaz, A.
  2001a, \apj, 563, L69

\bibitem[]{}Soker, N., Rappaport, S., \& Fregeau, J. 2001b, \apj, 563,
  L87
\bibitem[]{}Sweigart, A. V. 1997, \apj, 474, L23
\bibitem[]{}Tayler, R. J., \& Wood, P. R. 1975, MNRAS, 171, 467
\bibitem[]{}Tokunaga, A. T. 2000, in Allen's Astrophysical Quantities, 4th ed.,
ed. A. N. Cox (New York: Springer), 143
\bibitem[]{}VandenBerg, D. A., \& Faulkner, D. J. 1977, ApJ, 218, 415

\bibitem[]{}VandenBerg, D. A., Bergbusch, P. A., \& Dowler, P. D. 2006,
  \apjs, 162, 375
\bibitem[]{}van Loon, J. Th., Boyer, M. L., \& McDonald, I., 2008,
\apj, 680, L49
\bibitem[]{}van Loon, J. Th., Stanimirovi\'c, S., Evans, A., \& Muller,
  E. 2006, \mnras, 365, 1277
\bibitem[]{}Ventura, P., \& D'Antona, F. 2005, \aap, 439, 1075
\bibitem[]{}Vink, J. S., \& Cassisi, S. 2002, \aap, 392, 553
\bibitem[]{}Wang, Y.-M. 1998, in Cool Stars, Stellar Systems, and the
  Sun: Tenth Cambridge Workshop, ed. R. A. Donahue \&
  J. A. Bookbinder, ASP Conf. Ser. 154, 131
\bibitem[]{}Yong, H., Demarque, P., \& Yi, S. 2000, \apj, 539, 928
\bibitem[]{}Zarro, D. M., \& Zirin, H. 1986, \apj, 304, 365 
\bibitem[]{}Zirin, H. 1975, ApJ, 199, L63
\bibitem[]{}Zirin, H. 1982, \apj, 260, 655
\end{thebibliography}
\end{document}